\documentclass{JFM-FLM_Au}

\usepackage{url}

\usepackage{siunitx}
\usepackage[finalnew]{trackchanges}
\usepackage{soul}

\addeditor{SP}
\newcommand{\Ek}{\textit{Ek}}

\renewcommand{\vec}[1]{\boldsymbol{#1}}

\newcommand{\D}{\mathrm{D}}
\renewcommand{\d}{\mathrm{d}}

\newcommand{\tke}{\mathrm{TKE}}
\newcommand{\ske}{\mathrm{SKE}}

\lefttitle{S. Peng, S. Silvestri, and A. Bodner}
\righttitle{Turbulent ocean front under strain}

\title{Submesoscale and boundary layer turbulence under mesoscale forcing in the upper ocean}

\author{Shirui Peng\aff{1}, Simone Silvestri\aff{2,1} \and Abigail Bodner\aff{1,3}}

\affiliation{\aff{1}Department of Earth, Atmospheric and Planetary Sciences, Massachusetts Institute of Technology, Cambridge, MA 02139, USA
\aff{2}Department of Environment, Land and Infrastructure Engineering, Politecnico di Torino, Torino, Italy
\aff{3}Department of Electrical Engineering and Computer Science, Massachusetts Institute of Technology, Cambridge, MA 02139, USA}

\corresau{Abigail Bodner, abodner@mit.edu}

\begin{document}
\maketitle

\begin{abstract}
The interaction among quasi-geostrophic mesoscale eddies, submesoscale fronts, and boundary layer turbulence (BLT) is a central problem in upper ocean dynamics. We investigate these multiscale dynamics using a novel large-eddy simulation on a \qty{100}{\kilo\meter}-scale domain with meter-scale resolution. The simulation resolves BLT energized by uniform surface wind and convective forcing. A front interacts with BLT within a prescribed, spatially inhomogeneous mesoscale eddy field, representing a canonical eddy quadrupole. Using a triple flow decomposition, we analyze the dynamic coupling and kinetic energy budgets among the large-scale field, submesoscale field, and the resolved BLT. Our analysis reveals significant heterogeneity in the structure and intensity of submesoscales and BLT under varying mesoscale forcing. Turbulent kinetic energy and production rates can vary by an order of magnitude along the front, creating distinct turbulent hotspots whose locations are tied to the underlying large-scale flow. The region under stronger mesoscale convergence holds stronger horizontal and vertical geostrophic shear productions for BLT, and stronger self-production and BLT-destruction for submesoscales. In contrast, the region under dominant mesoscale divergence holds dramatic distortion of the front isotherm, along with dominant submesoscale vertical buoyancy production and self-destruction. Within this idealized framework, these results provide a controlled process-level characterization of how prescribed mesoscale heterogeneity modulates BLT and submesoscales in the ocean mixed layer, which can inform future parameterization developments.
\end{abstract}

\begin{keywords}
Authors should not enter keywords on the manuscript, as these must be chosen by the author during the online submission process and will then be added during the typesetting process (see \href{https://www.cambridge.org/core/services/aop-file-manager/file/61436b61ff7f3cfab749ce3a/JFM-Keywords-Sept-2021.pdf.}{Keyword PDF} for the full list).  Other classifications will be added at the same time.
\end{keywords}


\section{Introduction}
\label{sec:intro}
The upper ocean is a turbulent interface between the atmosphere and the deep ocean, a reservoir of dynamical processes across a vast range of scales \citep[e.g.,][]{ferrari_ocean_2009}. At the largest scales, energy is injected by atmospheric winds and buoyancy fluxes and is organized into basin-scale gyres and mesoscale eddies, with typical scales of \textit{O}(10-100)~km. A portion of the energy is transferred towards smaller scales, notably into a vibrant field of submesoscale fronts and filaments (\textit{O}(0.1-10)~km), which can involve both upscale and downscale transfers, before its ultimate dissipation by microscale turbulence \citep[e.g.,][]{mcwilliams_submesoscale_2016, taylor_submesoscale_2023}. The submesoscale is a dynamically critical regime where the influence of planetary rotation weakens, and ageostrophic, three-dimensional motions emerge, fundamentally controlling the vertical exchange of heat, carbon, and nutrients between the surface and the ocean interior \citep[e.g.,][]{levy_bringing_2012, su_ocean_2018, boyd_multi_2019}.

A central challenge in oceanic geophysical fluid dynamics is to understand and quantify the interaction between the largely balanced, quasi-geostrophic mesoscale flow and the fully three-dimensional, ageostrophic turbulence. Submesoscale interactions in the mixed layer (ML)--the near-surface layer that is actively homogenized by atmospheric forcing, and in particular submesoscale fronts--sharp, horizontal density gradients, serve as the natural laboratory for this problem. In the real ocean, ML temperature fronts can result from passing storm events, leaving abnormal horizontal ML gradients above a stratified interior. They can be energized by the strain field of the background mesoscale flow, which acts to sharpen the density gradients in a process known as frontogenesis \citep{hoskins_atmospheric_1972}. Realistic simulations also discovered that ML temperature fronts can be strengthened by a balance known as Turbulent Thermal Wind (TTW) \citep{gula_submesoscale_2014, crowe_evolution_2018}. The frontogenetic evolution further triggers various instabilities, including symmetric, baroclinic, and shear instabilities. Symmetric and shear instabilities can drive intense turbulence and vertical mixing that arrest the frontal collapse, while baroclinic instabilities (BI) release the potential energy (PE) in the front and restratify the ML \citep{capet_mesoscale_2008b, sullivan_frontogenesis_2018}. 

Ocean fronts also interact constantly with boundary layer turbulence (BLT) driven by surface wind shear, convection, or waves \citep{thorpe_turbulent_2005, mcwilliams_submesoscale_2017}. In particular, atmospheric wind and cooling can induce surface Ekman flow and convection, and their interactions with fronts can involve geostrophic shear production, Ekman buoyancy flux, or symmetric instability \citep[e.g.,][]{thomas_destruction_2005,mahadevan_rapid_2010, thomas_symmetric_2013,skyllingstad_baroclinic_2017, callies_baroclinic_2018, wenegrat_current_2023}. These interacting processes --- frontogenesis, mixed-layer instabilities (MLIs), restratification, and BLT --- together govern momentum fluxes and cross-scale energy pathways in the ML. In particular, MLIs in unstable fronts can generate submesoscale filaments and vortices while BLT simultaneously mixes and redistributes buoyancy, so the two regimes can coexist and mutually modulate each other \citep[e.g.,][]{hamlington_langmuir_2014, whitt_energetic_2017, sullivan_frontogenesis_2018, callies_baroclinic_2018, verma_interaction_2022}. Quantifying this coupling in a broader yet heterogeneous mesoscale background requires datasets that resolve processes from the mesoscale down to meter-scale BLT and diagnostics that isolate exchanges between submesoscale kinetic energy (SKE) and turbulent kinetic energy (TKE).

However, a persistent gap exists in modeling capabilities, leaving this multiscale interaction only partially resolved in existing model hierarchies. Global ocean general circulation models are advancing towards resolving the mesoscale but must still parameterize the effects of both submesoscale features and BLT \citep{ferrari_frontal_2011, bodner_modifying_2023, wagner_formulation_2024, silvestri_gpu_2025}. Regional hydrostatic models can successfully capture mesoscale-to-submesoscale transition. These simulations resolve how mesoscale strain generates sharp submesoscale fronts, in which many of the submesoscale features, especially those that develop in winter, are generated by MLIs \citep[e.g.][]{srinivasan_forward_2023,delpech_eddy_2024}. Yet, they lack the resolution and non-hydrostatic physics required to explicitly represent BLT, relying instead on parameterizations that cannot capture the fine-scale structure of vertical transport, energy dissipation and their interactions with the larger scales. On the other hand, process-oriented Large Eddy Simulations (LES) have provided invaluable insight on submesoscale-BLT dynamics, such as Langmuir-submesoscale interactions, turbulence-induced frontogenesis and frontal arrest, etc \citep[e.g.,][]{hamlington_langmuir_2014, skyllingstad_baroclinic_2017, sullivan_frontogenesis_2018, verma_interaction_2022, bodner_modifying_2023}. In particular, under wind forcing in \citet{skyllingstad_baroclinic_2017}, wind-aligned Ekman transport strongly disrupts baroclinic development, with a dominant instability mechanism combining ageostrophic shear instability of wind-driven flow with symmetric instability of the frontal geostrophic shear. At the filament scale in \citet{sullivan_frontogenesis_2018}, submesoscale filament frontogenesis driven by BLT is arrested within less than a day at a width of approximately 100 m by a horizontal shear instability of the sharpened filament, followed by slow frontal decay through further turbulent mixing. When convective turbulence is present, the submesoscale strengthens overall under surface cooling, with frontogenetic tendency increasing to counter enhanced horizontal diffusion by convection-induced turbulence, while the interscale transfer of energy from submesoscale velocity gradients to finescale turbulence accounts for up to half of the total finescale energy production \citep{verma_interaction_2022}. These models explicitly resolve BLT but have historically relied on a critical idealization: the assumption of a spatially absent or uniform background mesoscale strain field. 

The hydrostatic or uniform-strain assumption neglects the fundamental reality that these scale interactions are concurrent. While analytically and computationally convenient, idealizing the mesoscale forcing as uniform omits the intrinsic spatial heterogeneity of the real ocean and its effects that propagate onto the submesoscale to BLT transition. This simplification not only filters out the direct influence of mesoscale strain gradients but also neglects the critical role of mesoscale vorticity, which can hold wave modes and instabilities, such as vortex-Rossby waves (VRW) and inter-eddy edge waves, that can significantly alter frontal turbulence evolution \citep[e.g.,][]{mcwilliams_formal_2003,dritschel_multiple_2008}. In the real ocean, submesoscale fronts and eddies do not exist in isolation; they are continuously interacting with both the heterogeneous mesoscale background and BLT. This leaves a foundational question unanswered: how does the spatial heterogeneity of the mesoscale field modulate the structure, energetics, and turbulent properties along a submesoscale front? Answering this requires a simulation that can simultaneously resolve BLT while explicitly accounting for a non-uniform, large-scale mesoscale field.

\begin{figure}[pt]
    \centering
    \includegraphics[width=\linewidth]{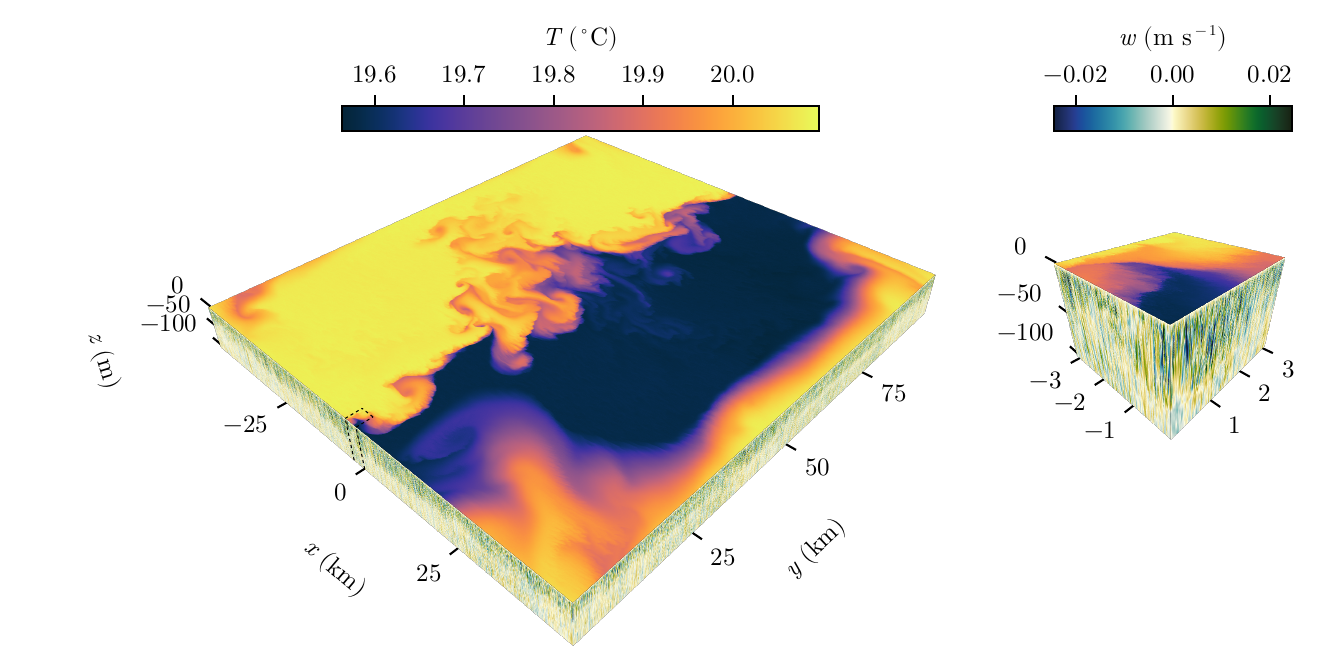}
    \caption{Visualization of surface temperature field $T$ and cross-section vertical velocity field $w$ in the computational domain above $z=\qty{-100}{\meter}$ at \qty{7.5}{\day}. Structures at multiple scales are developing in the visible horizontal plane (left), indicative of efficient energy transfer across scales. At the same time, 3D instability patterns and small-scale features are detectable in both horizontal and vertical cuts of the zoomed-in simulation domain (right), suggesting a forward turbulent cascade.\add[SP]{For clarity, the same zoomed-in domain is illustrated again in \protect{Fig.~\ref{fig:Tlast_app} (Appendix~\ref{app:zoomin})} with the same aspect ratio as the full domain.}}
    \label{fig:Tlast}
\end{figure}

In this study, we present results from a first-of-its-kind, large-domain LES that can resolve scales ranging from \qty{100}{\kilo\meter} down to single meters (Fig.\ref{fig:Tlast}). The defining feature of this study is the imposition of a prescribed, time-invariant, and spatially inhomogeneous background velocity field, $\vec{U}$, which represents a canonical mesoscale eddy quadrupole. The LES resolves the perturbation velocity field $\vec{u}$ evolving under the influence of this large-scale driver, as governed by the material derivative ${\D }/{\D t}={\p }/{\p t}+\left(\vec{u}+\vec{U}\right)\vec{\cdot \nabla}$. In our idealized framework, the mesoscale eddies are prescribed and stationary; therefore, we focus on mesoscale modulation of submesoscale and BLT energetics rather than on evolving mesoscale instabilities. Accordingly, the present framework isolates mesoscale modulation of perturbation/front dynamics but does not represent a prognostic, mutually advecting mesoscale eddy population. This configuration couples a BLT-resolving LES with a persistent, non-uniform mesoscale field, allowing us to directly investigate the dynamic response of the front to a spatially varying environment. To distinguish various components in this multiscale system, we develop and apply a triple flow decomposition to quantitatively isolate the energetics of the large-scale, the submesoscale turbulence, and BLT. We test the following hypotheses: (i) mesoscale convergence sharpens fronts and enhances geostrophic shear production of TKE, (ii) mesoscale divergence favors frontal distortion and stronger buoyancy-driven SKE pathways, and (iii) along-front heterogeneity in mesoscale forcing delays or accelerates local MLI growth relative to neighboring sectors.

By decomposing the flow into its prescribed mesoscale, along-front mean, filtered submesoscale, and BLT components, we quantify the dynamical coupling and kinetic energy budgets across scales. We demonstrate that the along-front structure of BLT is profoundly modulated by the mesoscale field, with local turbulent kinetic energy and production rates varying by an order of magnitude in response to heterogeneous background mesoscale and Ekman flow. This paper is structured as follows. §\ref{sec:setup} details the LES configuration, and §\ref{sec:decompose} the triple decomposition methodology. §\ref{sec:evolution} and §\ref{sec:results} presents the primary results, focusing on the spatially varying structure of the front, its associated turbulence, and the multiscale energy budgets. In §\ref{sec:discussion}, we discuss the implications for vertical transport and submesoscale parameterization and summarize our key conclusions.

\section{Numerical simulation}
\label{sec:setup}

The simulation presented in this paper is of a frontal spin down in a rotating, incompressible Boussinesq fluid. The governing equations for momentum, temperature, and mass are
\begin{subequations} \label{eq:bns}
  \begin{align}
 \frac{\D \vec{u}}{\D t}+f\hat{\vec{z}}\times \vec{u}&=- \vec{\nabla} \varphi+ b\hat{\vec{z}}+ \vec{F}_{\vec{u}} , \label{eq:momentum}\\
  \frac{\D T}{\D t}&=F_T, \label{eq:temperature}\\
 \vec{\nabla \cdot u}&=0, \label{eq:continuity}
\end{align}  
\end{subequations}
where 
\begin{equation}
    \frac{\D }{\D t}=\frac{\p }{\p t}+\left(\vec{u}+\vec{U}\right)\vec{\cdot \nabla}=\frac{\p }{\p t}+\left(u+U\right)\frac{\p}{\p x}+\left(v+V\right)\frac{\p}{\p y}+w\frac{\p}{\p z}
\end{equation}
is the Lagrangian derivative with $\vec{U}\equiv U \hat{\vec{x}}+V \hat{\vec{y}}$ a fixed background eddy strain field defined in Appendix~\ref{app:eddyv}, $f=\qty{1e-4}{\per\second}$ the Coriolis parameter at around \qty{45}{\degree} N, $\varphi$ the geopotential, and $b\equiv-g(\rho/\rho_0 -1)=\alpha g T$ is seawater buoyancy relative to a reference density $\rho_0=\qty{1020}{\kilo\gram\per\cubic\meter}$, $\alpha=\qty{2e-4}{\per\celsius}$ the thermal expansion coefficient, $g=\qty{9.81}{\meter\per\second\squared}$ the gravitational acceleration, and $T$ is temperature. Because $\vec{U}$ is prescribed, the model is one-way coupled with respect to mesoscale evolution; terms representing perturbation feedback onto a prognostic mesoscale state are not advanced. Velocity and temperature surface forcing terms are $\vec{F}_u$ and $F_T$, respectively. Note that density is only affected by temperature here for simplicity. The equation set \eqref{eq:bns} is integrated using Oceananigans, a Julia package for simulation of incompressible fluid flows in an oceanic context \citep{Oceananigans-overview-paper-2025}. The formulation is implicit LES (ILES): subfilter dissipation is provided numerically by the advection/discretization scheme rather than an explicit eddy-viscosity closure, so dissipation is interpreted through the residual/advection terms in the diagnosed budgets. The choice of no explicit closure has several advantages compared to typical explicit approaches, such as reduced computational cost, no parameter tuning, and higher effective resolution \citep{pressel, silvestri_new_2024} Pressure is obtained from the standard projection step with a distributed Poisson using discrete Fourier transforms. The domain is transposed when performing the three-dimensional DFT to enable strong scaling across the 1024-GPU decomposition used here. The grid size is $(N_x,N_y,N_z)=(20480,20480,224)$ and the simulation domain is $(x,y)\in [0, L_x)\times[0, L_y)$, where $L_x=L_y=\qty{100}{\kilo\meter}$, and $z\in [-L_z,0]$, where $L_z=\qty{252}{\meter}$. The grid spacing is $\Delta h=\qty{4.88}{\meter}$ and $\Delta z=\qty{1.125}{\meter}$ in the horizontal and vertical directions, respectively. We keep a uniform vertical grid primarily for numerical simplicity and GPU efficiency in this very large domain: a uniform direction allows using a three-dimensional discrete Fourier transform to solve the pressure Poisson equation rather than resorting to tri-diagonal implicit solve in the vertical. We therefore interpret the budgets as resolved-scale energetics and avoid claiming full resolution of all near-surface turbulent scales. The resolution is close to that used in \citet{hamlington_langmuir_2014} that resolves specific BLT-submesoscale interactions, although this model uses finite volume methods for horizontal derivatives instead of spectral derivatives as in \citet{hamlington_langmuir_2014}. This domain is partitioned into 1024 GPUs, which divide each horizontal dimension uniformly into 32 ranks. In comparison, a hydrostatic version of the simulation can run on 1 GPU with $\Delta h=\qty{156.25}{\meter}$ \citep{silvestri_gpu_2025}.

The domain is periodic in both horizontal directions. In the vertical, boundary conditions include momentum and temperature fluxes induced by a constant wind stress $\vec{\tau}$ and a uniform vertical temperature flux $J^T$ at the top surface as 
\begin{equation}\label{eq:top_cond}
  \vec{F}_{\vec{u}} = -\frac{\d \left[\vec{\tau} \delta(z)\right]}{\d z} = -\frac{\tau_w \left(\cos{\theta}\hat{\vec{x}}+\sin{\theta}\hat{\vec{y}}\right)\delta'(z)}{\rho_0}  ,\quad
  F_T=- \frac{\d \left[ J^T \delta(z)\right]}{\p z}=\frac{Q\delta'(z)}{\rho_0 c_p}, 
\end{equation}
where $|\vec{\tau}|=\tau_w=\qty{0.1}{\newton\per\meter\squared}$ is the wind stress magnitude, $\delta(z)$ a delta function concentrate at $z=0$, $\theta=\qty{30}{\degree}$ the wind angle, $Q=\qty{40}{\watt\per\meter\squared}$ the surface heat flux out of the ocean (positive for cooling), and $c_p=\qty{3995}{\joule\per\kelvin\per\kilo\gram}$ the heat capacity of seawater. At the surface, we impose an impermeable boundary ($w=0$) with prescribed momentum and heat fluxes. At $z=-L_z$, we impose free-slip and no-normal-flow conditions, ${\p u}/{\p z}={\p v}/{\p z}=w=0$, together with an initialized background temperature gradient $\alpha g{\p T}/{\p z}=0.1N^2_T$,
where $N^2_T=\qty{2e-4}{\per\second\squared}$ is a deep pycnocline water buoyancy frequency squared.

\begin{figure}[pt]
    \centering
    \includegraphics[width=0.9\linewidth]{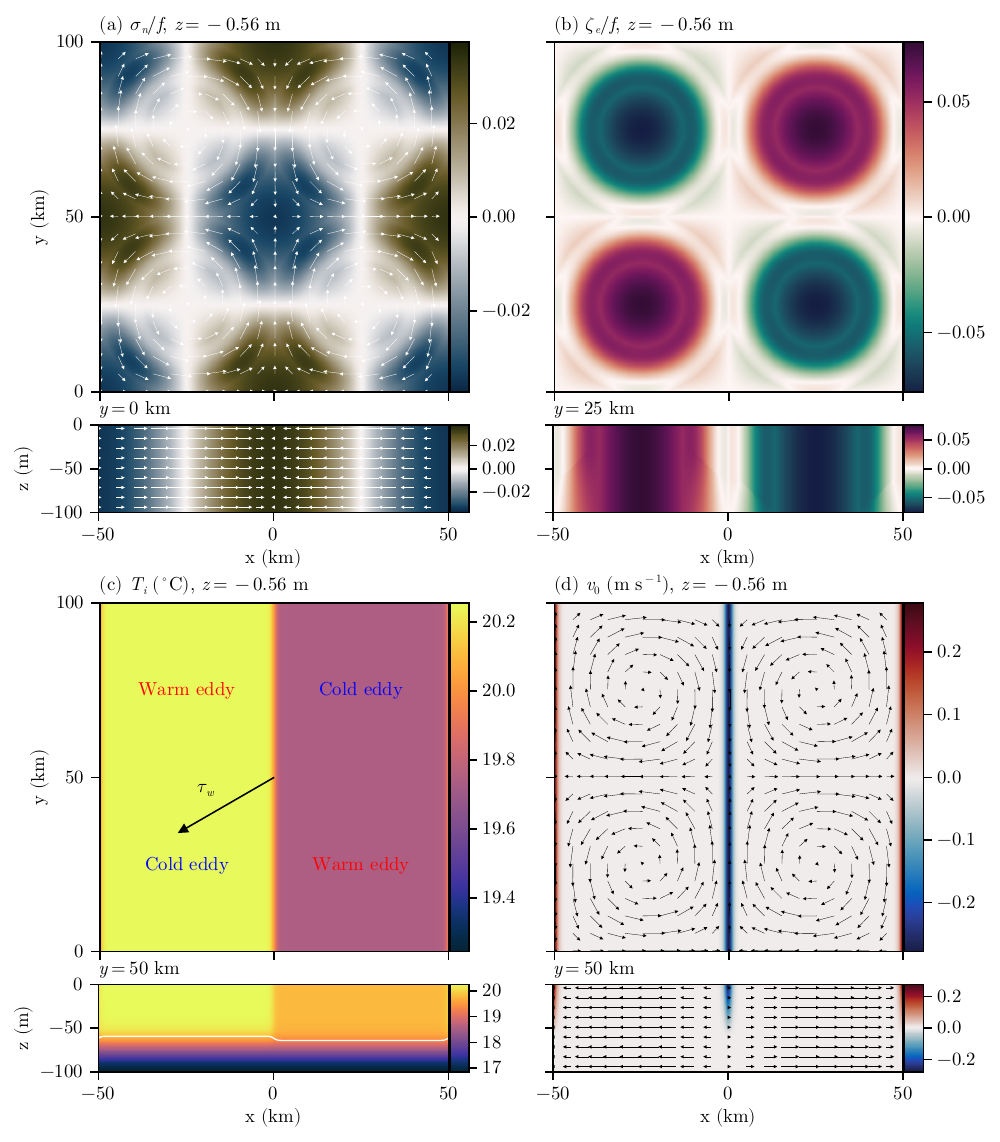}
    \caption{Snapshots of (a) background mesoscale strain rate $\sigma_n=-\frac{1}{2}(\p U/\p x-\p V/\p y)=-\p U/\p x$, (b) background mesoscale vorticity $\zeta_e=\p V/\p x-\p U/\p y$, (c) initial temperature $T_i$, and (d) initial jet velocity $v_0$. The $x-y$ slices are at $z=\qty{0.56}{\meter}$, and the $x-z$ slices are at $y=\qty{0}{\kilo\meter}$ for (a), $y=\qty{25}{\kilo\meter}$ for (b), and $y=\qty{50}{\kilo\meter}$ for (c,d). Arrows in (a,d) indicate velocity vectors of the eddy forcing. The black arrow in (a) indicates the wind direction. White lines in (a) show the initial mixed layer depth. Light rings in (b) that are roughly at a half radius point implies merged effects due to compactly positioned eddy quadruple.}
    \label{fig:T0v0}
\end{figure}

The prescribed background eddy velocity field $\vec{U}$ represents four stationary and near-barotropic mesoscale eddies arranged in a quadruple pattern (Fig.~\ref{fig:T0v0}ab). The setup is a combination of that in \citet{hamlington_langmuir_2014} without Stokes components and the one in \cite{thomas_friciton_2008} with strain. The eddies have a radius at $R=L_x/4=\qty{25}{\kilo\meter}$, a surface-level anomaly magnitude at $\Phi=\qty{0.01}{\meter}$, and a vertical temperature profile that is constant in the ML and decreases by $\Delta T^e = \qty{0.1}{\celsius}$ at the bottom. To represent the eddy effect on the isopycnal, the depth below which the temperature starts decreasing is slightly deeper(shallower) for warm(cold)-core eddies, with a difference of $\Delta m=\qty{30}{\meter}$. The eddy velocities consist of a dominant barotropic part corresponding to the surface-level anomaly and a mild baroclinic part related to the vertical temperature profile. The velocity has a maximum of \qty{0.058}{\meter\per\second} in the surface layer and decreases slightly at the bottom. This background field acts as an idealized mesoscale advection of temperature and momentum, while the eddy temperature $T_M$ is superimposed onto the background stratification as part of the initial condition. We prescribe the eddies as stationary over the analyzed time window to impose a controlled, spatially varying mesoscale strain pattern; we therefore do not represent mutual advection among mesoscale vortices, and we interpret all diagnostics within this idealized forcing context.

The initial state of the simulation is a temperature profile $T_i(x,y,z)=T_M(x,y,z)+T_0(x,z)$, and a jet $\vec{u}_i$=$v_0(x,z)\hat{\vec{y}}$ in thermal wind balance with $b_0=\alpha gT_0$ as
\begin{equation}
    v_0=\frac{1}{f}\int_{-\infty}^z \frac{\p b_0}{\p x}\, \d z.
\end{equation}
Here $T_M$ and $\vec{U}$ are constructed to be internally consistent with the intended thermal-wind scaling of the imposed mesoscale state (Appendix~\ref{app:eddyv}), while remaining an idealized analytic background rather than a fully prognostic balanced ocean vortex solution. We initialize $T_0$ as a two-front configuration (Fig.~\ref{fig:T0v0}bc) \citep{hamlington_langmuir_2014, verma_interaction_2022}. It consists of a weakly stratified surface layer above a strongly stratified thermocline and a moderately stratified lower layer. The analytical form of $v_0$ and $T_0$ is given in Appendix~\ref{app:initialTv}. The initial condition results in two fronts oriented in the $y$ direction, one located at $x=\qty{0}{\kilo\meter}$ and one at $x=\qty{50}{\kilo\meter}$, where the buoyancy gradient at each front has a magnitude of $M^2_0=\qty{5e-7}{\per\second\squared}$. Due to the periodicity of the domain and the imposed wind stress as described below, one front remains stable, whereas instabilities form along the other under both mesoscale forcing and Ekman buoyancy flux, an important mechanism for generating submesoscale activity \citep{thomas_destruction_2005, thomas_symmetric_2013, skyllingstad_baroclinic_2017}. The mean mixed-layer depth (MLD) is $m_0=\qty{60}{\meter}$ with an initial squared Brunt–V\"ais\"al\"a frequency $N^2_s=\qty{5e-7}{\per\second\squared}$, representative of the MLD in the western boundary current regions during October to December \citep{treguier_mixed-layer_2023}. We set a transition layer with a depth of $\Delta m^f=\qty{10}{\meter}$ between the ML base and the pycnocline. The pycnocline stratification is characterized by $N^2_T$ between \qtyrange{70}{100}{\meter} depth, below which the stratification is reduced by a factor of 10. For reference, the setup initializes a frontal strength $M^2/f^2=50$ with \textit{O}(1) maximum Rossby number.

We designed the stratification and surface forcing to ensure a representative and relatively stable evolution of the MLD \citep{legay_framework_2024}. Following the framework of \citet{legay_framework_2024}, we plot the initial horizontal distribution of four dimensionless parameters that characterize ML dynamics (Fig.~\ref{fig:mldass_d0} in Appendix~\ref{app:mldass}). The parameter $\lambda_s=-B_0 h/u_*^3$ represents the relative contribution of surface cooling and wind forcing, where $B_0=-\alpha g Q/(\rho_0 c_p)$ is the buoyancy flux, $h$ is the MLD computed based on a buoyancy difference of \qty{3e-4}{\meter\per\second\squared} from the surface \citep{griffies_omip_2016}, and $u_*=\sqrt{\tau_w/\rho_0}$ is the friction velocity. The two Richardson numbers, $R_h=(N_h h/u_*)^2$ and $R_h^*=(N_h h/w_*)^2$, describe the stability of the ML with respect to wind-driven turbulence and convective mixing, respectively. Here $N_h^2$ represents the stratification at the base of the ML, and $w_*=(-B_0 h)^{1/3}$ is the convective velocity scale. The last parameter, $f/N_h$, characterizes the influence of rotation relative to stratification. For reference, the ratios of $B_{\Ek}/B_0=25.5$ and $B_0/(f w_*^2)=1.75$ with $B_{\Ek}=\tau_w M^2_0/(\rho_0 f)$.

Red regions in Fig.~\ref{fig:mldass_d0} indicate parameter values that fall within the \qty{30}{\percent} highest-density contours in the global parameter space analyzed by \citet{legay_framework_2024} (see their Fig.~B1), which also broadly corresponds to conditions associated with stable MLD evolution (see their Fig.~2). Specifically, $\lambda_s\gtrsim10^{-0.6}\approx0.25$ indicates $B_0$ or convection, compared to $u_*$ or wind shear, as the driving force of MLD deepening (Fig.~\ref{fig:mldass_d0}a). And a pattern of $R_h\gtrsim10^3$ and $R_h^*\gtrsim10^{3.5}\approx 3\times 10^3$ implies stable ML stratification under the given surface forcing (Fig.~\ref{fig:mldass_d0}bd). Note that the effects of wind can be as strong as convection, even much stronger based on large $B_{\Ek}/B_0$ in the frontal region characterized by restratification and a shallower mixing depth. Although these thresholds come from 1D results and vary with horizontal heterogeneity \citep{legay_framework_2024}, we regard them as pragmatic dynamical bounds for the parameter choice. Notably, light color transitions in the plots coincide with the locations of submesoscale fronts and cold-core eddies, while warm-core eddies exhibit a weaker imprint. This difference likely arises because warm-core eddies deepen isopycnals, but their influence is masked by the stronger background stratification below the mean MLD.

\section{Multiscale energetics}
\label{sec:decompose}
The spatially inhomogeneous nature of the background flow $\vec{U}$ and the lack of time-averaged data necessitate a flow decomposition based on spatial filtering. We adopt the approach taken in \citet{johnson_modification_2024} for a multi-level decomposition for the total velocity field 
\begin{equation}
    \vec{u}_{total}=\vec{U}+{\vec{u}}=\vec{U}+\overline{\vec{u}}+\vec{u}'=\vec{U}+{\vec{u}^a}+{\vec{u}^s}+\vec{u}',
\end{equation}
where $\overline{\vec{u}}$ represents a 2D horizontal Gaussian filter with a length scale of \qty{300}{\meter} permitted as submesoscale \citep{hamlington_langmuir_2014,bodner_breakdown_2020}, and $\vec{u}'$ are the finer-than-submesoscale fluctuations, representing BLT. When using the filtering for subdomains, we apply reflective boundary conditions for non-periodic boundaries and focus on results away from the boundaries. The next decomposition separate the frontal flow from the wind-driven Ekman flow by an along-front average in the $y$ dimension $\langle{\overline{\vec{u}}}\rangle={\vec{u}^a}=\overline{\vec{u}}-{\vec{u}^s}$ \citep{johnson_modification_2024}. Note that this along-front average still contains a mean frontal component besides the Ekman flow. In this framework, we will have advection of the large-scale along-front mean flow ${\vec{u}^a}$, despite suppressing that for the barotropic eddies in equation~\eqref{eq:bns}, with the assumption being that the mesoscale eddies are stationary within the time frame relevant for the simulation physics. For any variable $\phi$, we denote $\overline{\vec{u}'\phi'}=\overline{\vec{u}\phi}-\overline{\vec{u}}\overline{\phi}$ \citep{germano_turbulence_1992}. We define the submesoscale kinetic energy, $\ske\equiv \frac{1}{2}\vec{u}^s \vec{\cdot} \vec{u}^s$ \citep{verma_interaction_2022}, and the turbulent kinetic energy (TKE) as $\tke\equiv\frac{1}{2}(\overline{u'^2}+\overline{v'^2}+\overline{w'^2})$. That is, we filter velocity first and then form TKE from the residual velocity squares; we do not apply a spatial filter directly to an already-constructed TKE field. This also differs from the formulation in \protect{\citet{verma_interaction_2022}} that squares the residual velocities.


\section{Unstable front develops heterogeneous structures}\label{sec:evolution}
\begin{figure}[pt]
    \centering
    \includegraphics[width=0.9\linewidth]{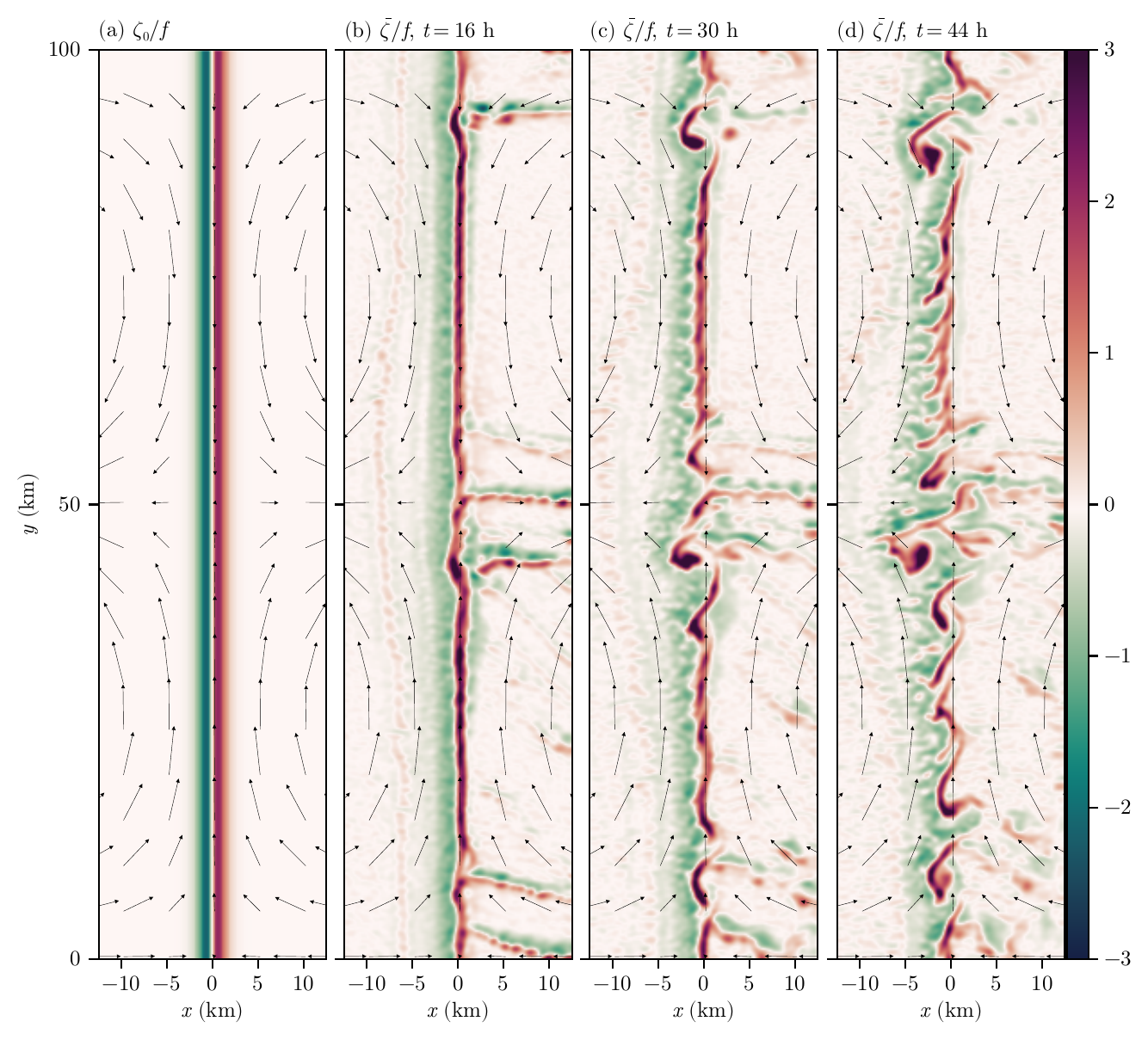}
    \caption{Snapshots of normalized surface vertical vorticity at (a) initialization, (b) 16~h, (c) 30~h, and (d) 44~h. The initial vorticity in (a) is calculated on a coarser grid with a horizontal spacing of \qty{156.25}{\meter}. This grid is enough to resolve the initial frontal jet and is created with 1 GPU. We do not need filtering since no turbulence is initialized. While those in other panels are calculated on the meter-scale grid after a \qty{300}{\meter}-Gaussian kernel smoothing. Arrows indicate velocity vectors of the eddy forcing. }
    \label{fig:curl}
\end{figure}

In this study, we focus on a small subset of the simulation output, the unstable frontal region before its fragmentation under submesoscale meandering and BLT, to ensure the analysis is computationally trackable. We explore the evolution of the simulation at later times in a follow up study. This region, between \qtyrange{-12.5}{12.5}{\kilo\meter} in $x$, shows rich multiscale evolution in the along-front direction (Fig~\ref{fig:curl}). We first characterize the flow in a qualitative overview of the front's structural evolution within the varying mesoscale eddy field. 

The evolution of the surface vertical vorticity field highlights the frontal response to the heterogeneous mesoscale flow (Fig~\ref{fig:curl}). Initially, the front is characterized by two parallel bands of cyclonic ($\zeta_0/f>0$) and anticyclonic ($\zeta_0/f<0$) vorticity of equal magnitude, consistent with the thermal wind balance of the initial state (Fig~\ref{fig:curl}a). By $t=\qty{16}{\hour}$, the cyclonic band sharpens and intensifies with significant local Rossby number $\overline{\zeta}/f=\textit{O}(1)$ (Fig~\ref{fig:curl}b). Concurrently, two groups of enhanced vorticity band emerge for $x\ge0$, one around $y=0$ and the other around $y=\qty{50}{\kilo\meter}$, corresponding to regions of background convergence and divergence. By $t=\qty{30}{\hour}$, adjacent to these bands significant curvature emerges on the along-front cyclonic band, and the two features slightly downstream of the strain extremes undergo severe meander (Fig~\ref{fig:curl}c), signifying the growth of frontal BI. By $t=\qty{44}{\hour}$, the front has fragmented into a complex field of submesoscale eddies and filaments, indicating that instabilities have grown to a developed stage (Fig~\ref{fig:curl}d). Consistent with the background mesoscale forcing, the submesoscale features are more spread out in the diverging zone ($y=\qty{50}{\kilo\meter}$) than they are in the converging zone ($y=\qtylist{0;100}{\kilo\meter}$). Similar evolution also shows up in the sufrace divergence plot (Fig~\ref{fig:div} in Appendix~\ref{app:div}). And we use spectra profiles at $t=\qty{44}{\hour}$ to clarify the multiscale resolution of the model and how TKE appears at scales comparable to submesoscale fields (Fig~\ref{fig:spec} in Appendix~\ref{app:spec}).

\begin{figure}[pt]
    \centering
    \includegraphics[width=0.9\linewidth]{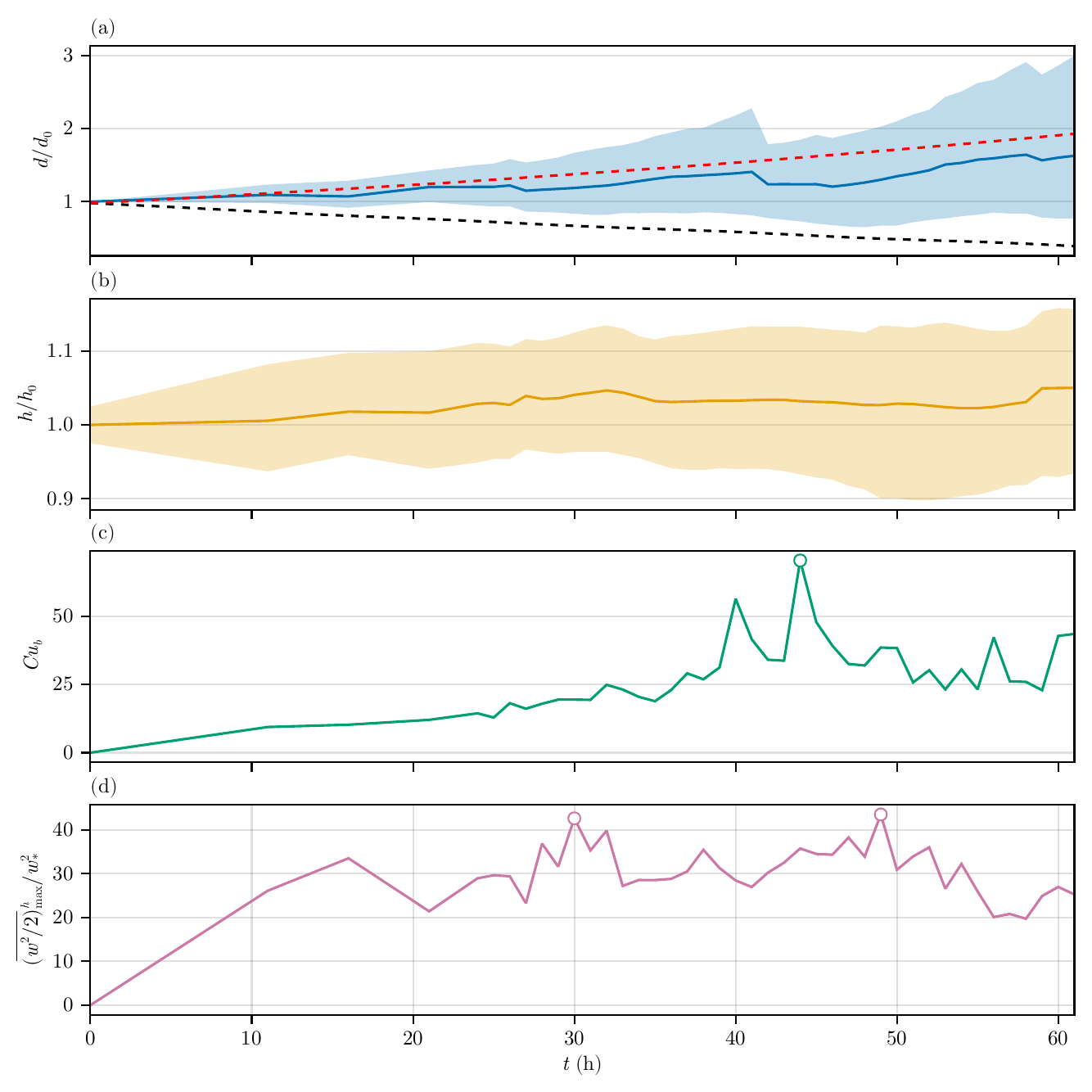}
    \caption{Time evolution for (a) mean frontal width $d$ normalized by $d_0=\qty{2}{\kilo\meter}$, (b) mean mixed layer depth $h$ in the front region normalized by $h_0=\qty{60}{\meter}$, (c) bulk curvature number $Cu_b$ along the central front isotherm, and (d) mixed-layer-averaged vertical kinetic energy maximum $\overline{w^2/2}^h_{\max}$ in the front region normalized by $w_*^2$. Shadings in (a)(b) illustrate the 10-th and 90-th percentile range, while circles in (c)(d) local peaks. Black and red curves in (a) are predictions from uniform strain frontogenesis theory \citep{shakespeare_generalized_2013}, with $\sigma_n=\pm 0.03 f$ for black and red, respectively. The bulk curvature number is the maximum along-front local curvature number $Cu\equiv 2u_g \kappa/f$, with $u_g=\alpha g \Delta T h/fd$ the geostrophic velocity scale and $\kappa$ the geometric curvature. Only solid lines of $d$ and $h$ are averaged over the whole frontal domain.}
    \label{fig:tsfront}
\end{figure}

The qualitative evolution discussed above is quantified by tracking several frontal properties over time (Fig.~\ref{fig:tsfront}). We use three surface temperature contours to define a consistent frontal region (Fig.~\ref{fig:ufront30h}a). Specifically, we first use the marching squares algorithm to find isolines of the spatially-smoothed surface temperature $\overline{T}_{\text{surf}}$. We then select those isotherms with small $x$-ranges $x_{\max}-x_{\min}<\qty{12.5}{\kilo\meter}$ and large $y$-ranges $y_{\max}-y_{\min}>\qty{90}{\kilo\meter}$ as frontal contour candidates. We identify the front's centre contour as the isoline of the mean temperature $\overline{T}_{\text{cfront}}$ in this candidate set. Two additional isotherms $\overline{T}_{\text{cfront}}\pm \qty{0.194}{\celsius}$ are then selected as frontal boundaries such that the initial frontal width is $\qty{2}{\kilo\meter}$. We present analysis on the frontal width, MLD, bulk curvature number, and vertical kinetic energy. The bulk curvature number $Cu_b$ is defined as the maximum value of the along-front curvature normalized by the geostrophic balance scale \citep{shakespeare_curved_2016}, giving the curvature number $Cu\equiv 2u_g \kappa/f$, with $u_g=\alpha g \Delta T h/fd$ the geostrophic velocity scale and $\kappa$ the geometric curvature.

Both the along-front mean and variation of normalized frontal width $d/d_0$ and MLD $h/h_0$ increase with time under yet different patterns and evolution dynamics (Fig.~\ref{fig:tsfront}ab). We define $d$ along the centre frontal contour as the sum of the smallest distances to the two frontal boundaries. Here we compute $h$ based on a density difference of \qty{0.09}{\kilo\gram\per\cubic\meter} from $z=\qty{-10}{\meter}$. This threshold is three times the conventional choice of \qty{0.03}{\kilo\gram\per\cubic\meter} \citep{treguier_mixed-layer_2023}, which can better capture the base of the ML under active multiscale processes \citep{epke_overturning_2025}. The 10-th and 90-th percentile bounds of $d/d_0$ are strongly \change[SP]{coupled to}{correlated with} the heterogeneous strain field, as will be confirmed in the along-front analysis later. Compared to the frontogenesis theory proposed by \citet{shakespeare_generalized_2013} based on a constant strain $\sigma/f=\pm0.03$ close to the prescribed mesoscale extremes here, the 10-th percentile of $d/d_0$ closely follows the prediction for converging strain up to about \qty{50}{\hour}, while the 90-th percentile increases more rapidly than the theory with a constant diverging strain. This asymmetry suggests that processes apart from \change[SP]{strain}{mesoscale forcing, e.g. instability-induced submesoscale meandering,} could be contributing significantly to widen the front. Here these dynamics could involve BLT-induced frontal arrest and diffusion and growth of submesoscale BI (Fig.~\ref{fig:ufront30h}).

For the normalized MLD $h/h_0$, the 10-th and 90-th percentiles likely reflect compound effects from initialization across-front differences, restratification, and mixing (Fig.~\ref{fig:tsfront}a,~\ref{fig:ufront30h}ab). In particular, the finite percentile range at $t=0$ result from shallower versus deeper initialized MLD $h$ for warmer and colder ML, respectively (Fig.~\ref{fig:T0v0}c). The time evolution of $h/h_0$ reveals a spatially and temporally variable interplay between BLT-driven deepening and submesoscale restratification \citep[e.g.][]{capet_mesoscale_2008b, mcwilliams_submesoscale_2016, verma_interaction_2022}. The asymmetry here is weaker yet shows a slightly stronger increase in the 90-th percentile in the earlier stage, in contrast to a stronger 10-th percentile variability in the later stage after the bulk curvature number peak. This is consistent with faster BLT-deepening and slower submesoscale-restratification. \change[SP]{We further provide direct, localized evidence for this coupling in the subsequent vertical cross-section analysis.}{In the subsequent vertical cross-section analysis, we provide qualitative local context for this coupling rather than a standalone restratification diagnosis.} Comparing the time series of $d/d_0$ and $h/h_0$ also indicates a more significant \change[SP]{strain}{mesoscale} impact on the former based on the frontogenesis analysis. 

The bulk curvature number $Cu_b$ grows approximately linearly until about $t\approx\qty{30}{\hour}$ (Fig.\ref{fig:tsfront}c). After this point, its growth accelerates dramatically, peaking around $t=\qty{44}{\hour}$ as the front visibly fragments (Fig.\ref{fig:curl}c). The MLD-averaged vertical kinetic energy, a proxy for BLT intensity, exhibits a primary peak around $t=\qty{30}{\hour}$ (Fig.\ref{fig:tsfront}d). This timing could suggest a transition from BLT-dominant early evolution to submesoscale-dominant instabilities which leads to a secondary peak in vertical kinetic energy around $t=\qty{48}{\hour}$ . We focus this evolution analysis prior to $t=\qty{61}{\hour}$, after which the frontal fragmentation makes a coherent front-following analysis less reliable.

\section{Multiscale interactions along the unstable front with changing strain}
\label{sec:results}
After identifying a dynamically representative time, $t=\qty{30}{\hour}$, we transition to a quantitative analysis in a front-following coordinate system. In this coordinate, we examine how properties such as the MLD, frontal width and curvature, TKE, and SKE vary along the frontal coordinate relative to the local mesoscale strain. This analysis is followed by a systematic comparison of various TKE and SKE budget terms across four vertical slices along the front. \remove[SP]{And the last subsection present a description of a secondary mesoscale-Ekman effect that we find to contribute to the early along-front heterogeneity.}

\subsection{Mesoscale forcing induces non-uniform turbulent coupling along the front}
\label{subsec:afront}
\begin{figure}[pt]
  \centering
  \includegraphics[width=0.9\linewidth]{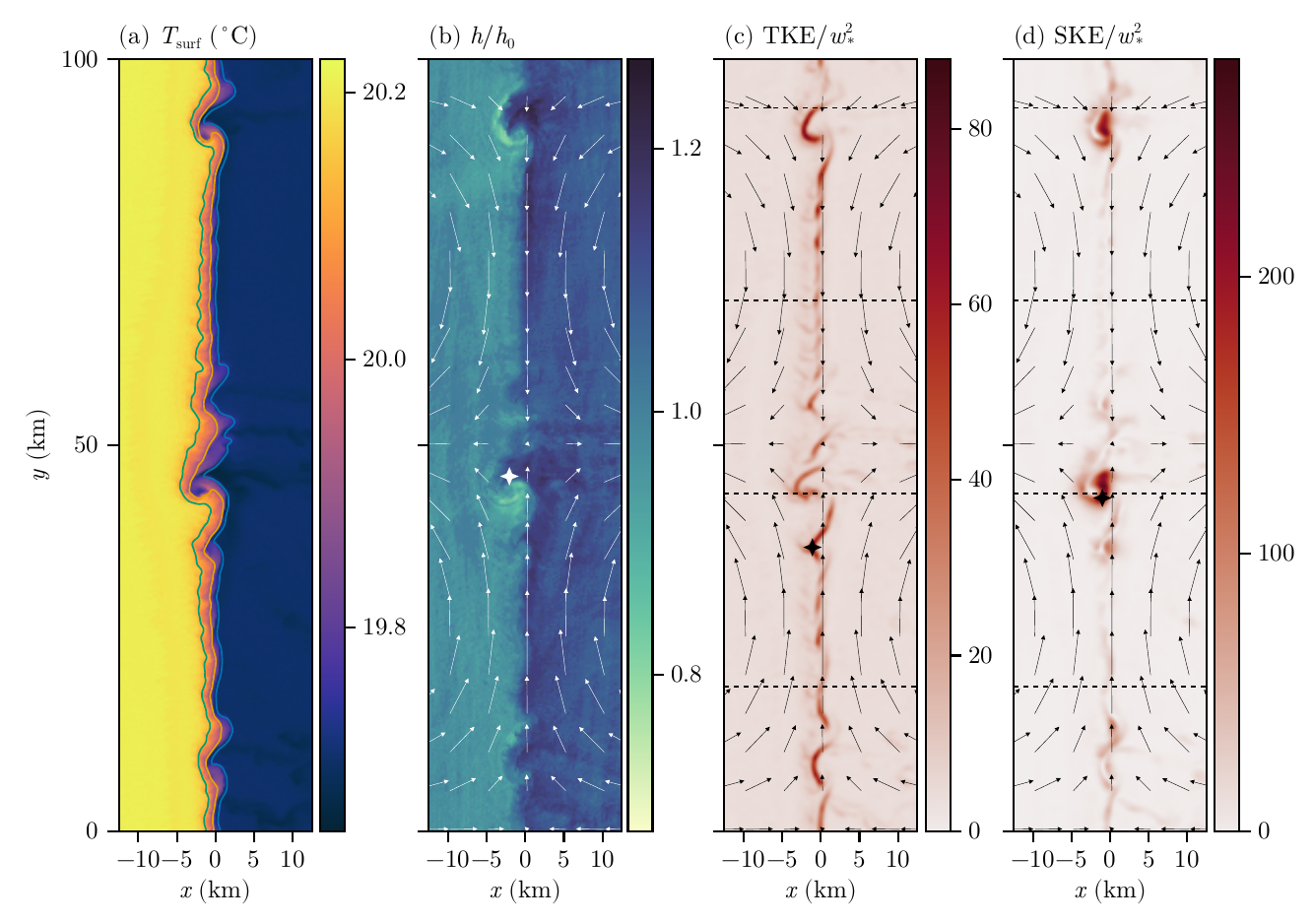}
\caption{Horizontal slice of (a) surface temperature $T_{\text{surf}}$, (b) normalized mixed layer depth $h/h_0$, (c) normalized turbulent kinetic energy $\text{TKE}/w_*^2$, and (d) normalized submesoscale kinetic energy $\text{SKE}/w_*^2$ at 30~h. The vertical level is $z=\qty{-0.56}{\meter}$ for TKE and $z=\qty{-2.81}{\meter}$ for SKE, which is based on where the maximum is located. The center contour in (a) corresponds to \qty{19.93}{\celsius}, and the two boundary contours are offset by \qty{0.194}{\celsius} which gives an initial frontal width of \qty{2}{\kilo\meter}. Arrows in (b)-(d) shows the velocity vector field that induces the strain, while white and black stars the maximum of $h$-averaged vertical kinetic energy, near-surface TKE ($z>\qty{-4.5}{\meter}$), and near-surface SKE ($z>\qty{-4.5}{\meter}$), respectively. Four dashed lines in (c,d) show slice locations along $y=\qtylist{93.75;68.75;43.75;18.75}{\kilo\meter}$ to be discussed in §\ref{subsec:budgets}.}
  \label{fig:ufront30h}
\end{figure}

\begin{figure}[pt]
  \centering
  \includegraphics[width=0.9\linewidth]{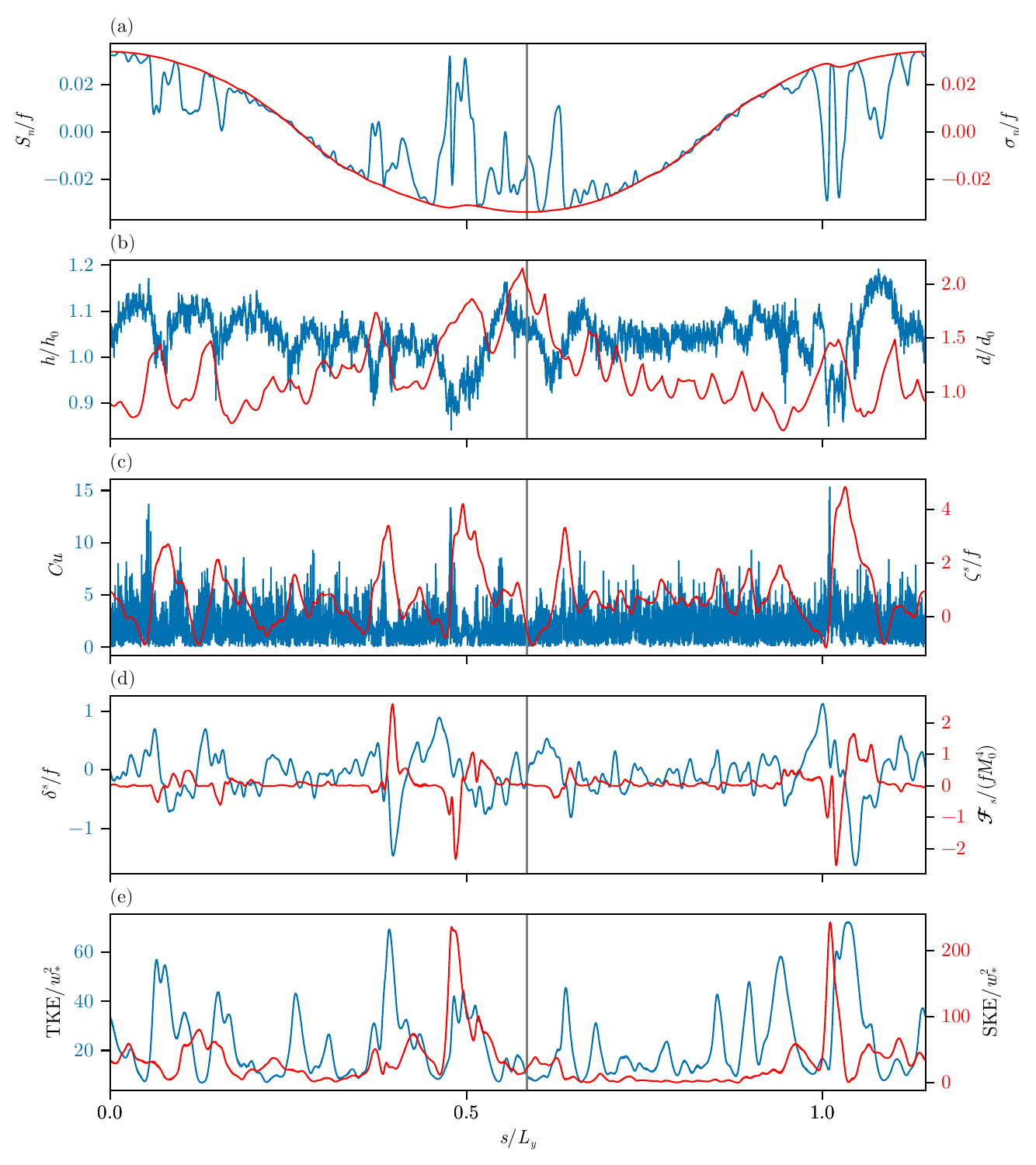}
\caption{Along-front profiles of (a) surface frontal normal strain $S_n/f$ and background strain $\sigma_n/f$, (b) surface frontal width $d/d_0$ and mixed layer depth $h/h_0$, (c) local curvature number $Cu$ and submesoscale vorticity ${\zeta}^s/f=({v}^s_x-{u}^s_y)/f$, (d) submesoscale divergence ${\delta}^s/f=({u}^s_x+{v}^s_y)/f$ and frontogenetic tendency $\mathcal{F}_s=-({b^s_x}^2 {u}^s_x+{b^s_y}^2 {v}^s_y)-{b}^s_x {b}^s_y ({u}^s_y+{v}^s_x)$ normalized by $f M_0^4$, and (e) turbulent kinetic energy $\text{TKE}/w_*^2$ and submesoscale kinetic energy $\text{SKE}/w_*^2$. All taken at 30~h and on the center contour of Fig.~\ref{fig:ufront30h}a. All submesoscale variables are computed on $z=\qty{-2.81}{\meter}$. The grey vertical line denotes where $\sigma_n/f$ reaches the strongest divergence (minimum).}
  \label{fig:alongfront30h}
\end{figure}

Based on the temporal evolution, we select $t=\qty{30}{\hour}$ for a detailed investigation. It represents a mature frontal state with strong turbulent activity, immediately prior to the dominance of extreme frontal curvature and further developed instabilities. Horizontal slices at this time reveal an unstable front around $x=0$ holding various emerging perturbations (Fig.~\ref{fig:ufront30h}). Signatures of emerging meanders are visible in the surface temperature field $T_{\text{surf}}$, particularly in the adjacent downstream regions of strong mesoscale strain (Fig.~\ref{fig:ufront30h}a). The MLD $h/h_0$ remains generally deeper on the cold, dense side of the front, though localized shallowing is evident near the emerging meanders, suggesting restratification effect by submesoscale BI (Fig.~\ref{fig:ufront30h}b). Hotspots of normalized near-surface TKE are concentrated along the front (Fig.~\ref{fig:ufront30h}c), with its maximum located slightly further downstream from the ML-averaged vertical kinetic energy maximum (black star in Fig.~\ref{fig:ufront30h}c vs. white star in Fig.~\ref{fig:ufront30h}b). This difference is consistent with possibly deeper MLD at the TKE maximum location, compared to shallower MLD at the maximum location of ML-averaged $\frac{1}{2}w^2$ under submesoscale restratification. In fact, the SKE maximum is nearly co-located with that of ML-averaged vertical kinetic energy. More generally, normalized near-surface SKE is concentrated into two strongest meandering modes near $y=\qtylist{44;93}{\kilo\meter}$, indicative of two emerging ML eddies due to submesoscale BI (Fig.~\ref{fig:ufront30h}d). Note that due to the removal of the along-front average in the multiscale decomposition §\ref{sec:decompose} the SKE field in Fig.~\ref{fig:ufront30h} reflects submesoscale instabilities that deviate from the mean frontal structure, such as BI and smaller fronts and filaments. 

To investigate the influence of the mesoscale \change{strain field}{forcing}, we project various quantities onto a coordinate system defined by the frontal arc length, $s$ (Fig~\ref{fig:alongfront30h}). We first resample the centre frontal contour into $N=10240$ points $\{\vec{r}_i=(x_i, y_i)\}_{i=1}^N$ and compute cumulative sums of their interval lengths to get $s$. The normalized coordinate $s/L_y$ is helpful to identify along-front locations relative to $y$. We define the instantaneous strain rate along the frontal coordinate normal $\vec{n}=(n_x,n_y)\propto \d^2 \vec{r}/ds^2$ as $S_n=-[n_x^2 \partial U/\partial x + n_y^2 \partial V/\partial y + n_x n_y  (\partial U/\partial y + \partial V/\partial x)]$. For other variables of interest, we interpolate the corresponding horizontal field onto $\{\vec{r}_i\}_{i=1}^N$ using nearest-neighbor and periodic boundary condition. 

The instantaneous normal strain $S_n/f$ generally overlaps with the time-invariant normal strain $\sigma_n/f=\frac{1}{2}(\p U/\p x-\p V/\p y)/f$ (Fig~\ref{fig:alongfront30h}a blue vs. orange curves). The MLD $h/h_0$ is reasonably noisy since its computation is based on the raw temperature field (Fig~\ref{fig:alongfront30h}b). It tends shallower when there is significant strain perturbation around $s/L_y=\qtylist{0.5;1.0}{}$, where submesoscale BI are located. The frontal width $d/d_0$ shows a strong anti-correlation with $\sigma_n/f$: it is maximized with strain minimum and minimized with strain maximum, consistent with the front being physically stretched and compressed (Fig~\ref{fig:alongfront30h}b, also~\ref{fig:ufront30h}a). Peaks in along-front local curvature number $Cu=2u_g \kappa/f$ are co-located with enhanced submesoscale vorticity $\zeta^s=({v}^s_x-{u}^s_y)/f$(Fig~\ref{fig:alongfront30h}c). Submesoscale divergence $\delta^s=({u}^s_x+{v}^s_y)/f$ reaches local minimum roughly with the local maximum of frontogenetic tendency $\mathcal{F}_s=-({b^s_x}^2 {u}^s_x+{b^s_y}^2 {v}^s_y)-{b}^s_x {b}^s_y ({u}^s_y+{v}^s_x)$, although the former shows higher overall variability (Fig~\ref{fig:alongfront30h}d). Similarly, TKE appear as random pulses with enhanced magnitude near strain extremes, while SKE is dominated by two maxima aligning with background strain extrema (Fig~\ref{fig:alongfront30h}e). Two SKE peaks match two dips in $\mathcal{F}_s$, dynamically consistent with finite amplitude growth of BI. Although this along-front correlation analysis does not necessarily transfer to causal dynamics, we can better understand the local and instantaneous dynamics by examining contributions to TKE and SKE budgets across a few vertical sections with distinct background strain. \add[SP]{In this context, the section-based diagnostics are interpreted as reflecting both mesoscale modulation and intrinsic submesoscale variability, rather than as a strict causal separation between the two.}

\begin{figure}[pt]
  \centering
  \includegraphics[width=\linewidth]{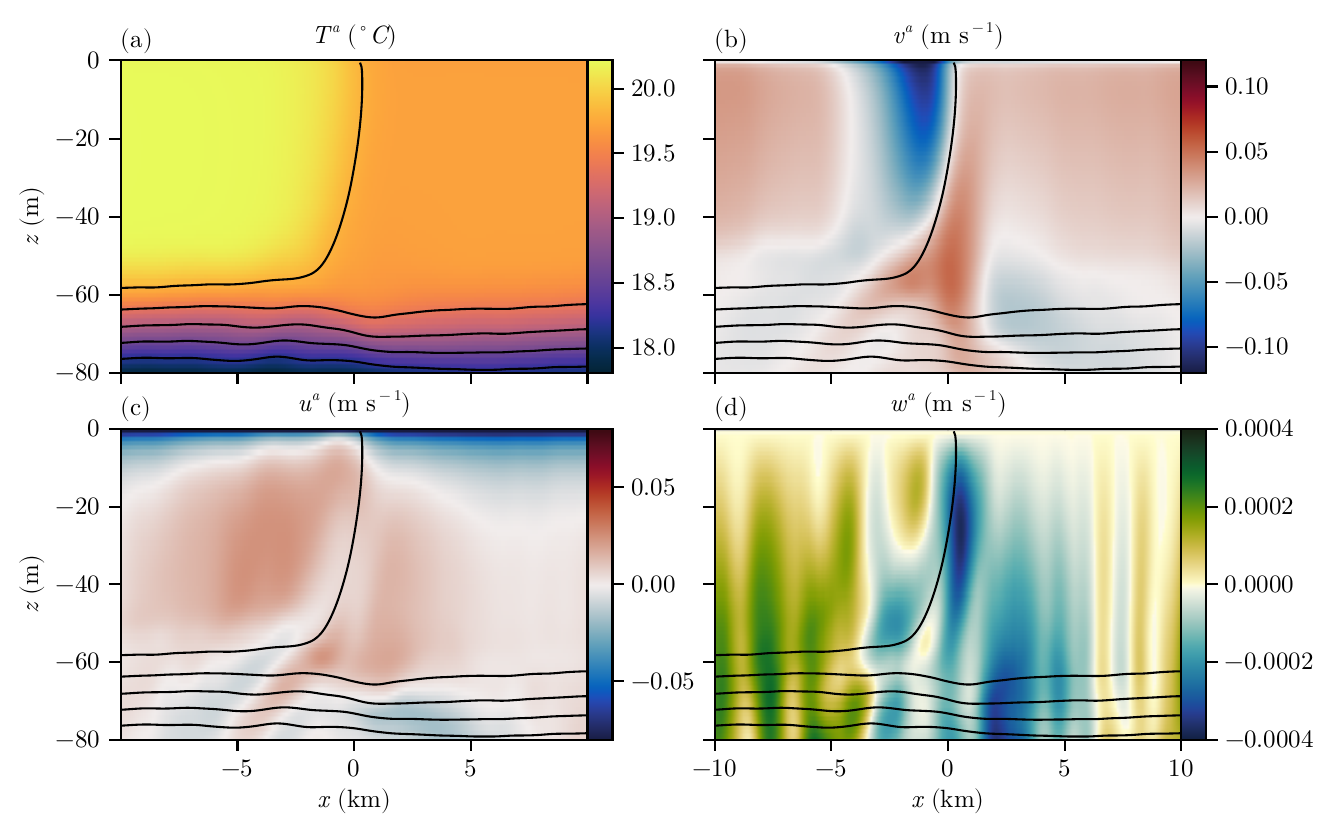}
\caption{Along-front mean fields of (a) temperature $T^a$, (b) along-front velocity $v^a$, (c) across-front velocity $u^a$, and (d) vertical velocity $w^a$ at \qty{30}{\hour}. \add[SP]{ Solid black lines indicate isotherms based on $T^a$.}}
  \label{fig:affields}
\end{figure}

Since the calculation of the SKE budget involves along-front decomposition, for computational efficiency we first divide \qty{100}{\kilo\meter} along $y$ into 16 same-size chunks. A sensitivity test using 4 chunks gives visually the same mean fields. Then we compute the along-front averages for each chunk and take the overall average at the end as the final result. The resulting mean fields show reasonable smoothed profiles highlighting the frontal and Ekman flows in $v^a$ and $u^a$, respectively (Fig.~\ref{fig:affields}bc blue colors). A mean secondary vertical flow $w^a$ is also obvious around the mean temperature front $T^a$ (Fig.~\ref{fig:affields}ad), likely an averaged ageostrophic response to strain and BLT \citep{shakespeare_generalized_2013, crowe_evolution_2018}. To eliminate boundary effects in filtering, we only average the \qty{5}{\kilo\meter} range around the center for each chunk, giving \qty{90}{\percent} along-front coverage.

\subsection{Mesoscale stirs heterogeneous energy distribution and conversion}
\label{subsec:budgets}

To obtain the evolution equation for $\ske\equiv \frac{1}{2}\vec{u}^s \vec{\cdot} \vec{u}^s$, we start from the momentum equations governing the filtered velocity $\overline{\vec{u}}$ \citep{verma_interaction_2022}
\begin{equation} \label{eq:ubar}
 \frac{\p \overline{u}_i}{\p t}+\frac{\p \overline{u}_i \left(\overline{u}_j+\overline{U}_j\right)}{\p x_j}+\epsilon_{i3k}f
\overline{u}_k=- \overline{\varphi}_j+ \alpha g \overline{T}\delta_{i3}+ \overline{F}_{u_i}-\frac{\p \tau_{ij}}{\p x_j} , 
\end{equation}
where $\tau_{ij}=\overline{u'_i{u}'_{total,j}}=\overline{u_i \left(u_j+U_j\right)}-\overline{u}_i \left(\overline{u}_j+\overline{U}_j\right)\approx\overline{u_i u_j}-\overline{u}_i \overline{u}_j$ is BLT residual stress. Subtracting the along-front average from \eqref{eq:ubar} and taking the dot product with $\vec{u}^s$ gives
\begin{equation}\label{eq:ske}
    \frac{\p \ske}{\p t}=P^a+P^s+P'+B^s+A^s,
\end{equation}
where 
\begin{subequations}
    \begin{align}
        P^a&=-\left[\vec{u}^s(\vec{u}^s+\vec{U})\right]\vec{: \nabla u}^a,\\
        P^s&=-\langle \vec{u}^s (\vec{u}^s+\vec{U})\rangle\vec{:\nabla u}^s,\\
        P' &= \left(\overline{\vec{u}'\vec{u}'_{total}}-\langle \overline{\vec{u}'\vec{u}'_{total}} \rangle\right)\vec{:\nabla u}^s,\\
        A^s &= -\vec{\nabla \cdot}\left\{\overline{\vec{u}}_{total} \ske + \vec{u}^s \varphi^s + \vec{u}^s \vec{\cdot} \left[\left(\overline{\vec{u}'\vec{u}'_{total}}-\langle \overline{\vec{u}'\vec{u}'_{total}} \rangle\right) - \langle \vec{u}^s (\vec{u}^s+\vec{U})\rangle\right]\right\}
    \end{align}
\end{subequations}
are the along-front mean, submesoscale, and BLT productions with $\vec{ab}\vec{: \nabla u}=a_i b_j u_{i,j}$, and advections of $\ske$ by total velocity and of submesoscale geopotential and residual stresses by $\vec{u}^s$. $B^s={w^s b^s}$ is the submesoscale buoyancy production. Importantly, the term $B^s$ captures the transfer of frontal PE reservoir to SKE through BI \citep{verma_interaction_2022}, and its average are central to ML parameterizations in large-scale models \citep{fox-kemper_parameterization_2008, bodner_modifying_2023}. The evolution equation for TKE is
\begin{equation}\label{eq:tke}
   \frac{\p \tke}{\p t}=P_H+P_V+B+A,
\end{equation}
where the horizontal and vertical productions are \citep{germano_turbulence_1992}
\begin{subequations}
    \begin{align}
        P_H&=-\overline{u'u'_{total}}\frac{\p \overline{u}}{\p x}-\overline{u'v'_{total}}\frac{\p {u}^s}{\p y}-\overline{v'u'_{total}}\frac{\p \overline{v}}{\p x}-\overline{v'v'_{total}}\frac{\p {v}^s}{\p y},\\
        P_V&=-\overline{u'w'}\frac{\p \overline{u}}{\p z}-\overline{v'w'}\frac{\p \overline{v}}{\p z}-\overline{w'^2}\frac{\p \overline{w}}{\p z}-\overline{w'u'_{total}}\frac{\p \overline{w}}{\p x}-\overline{w'v'_{total}}\frac{\p {w}^s}{\p y},
    \end{align}
\end{subequations}
$B=\overline{w'b'}$ is buoyancy production, and $A$ is advection and dissipation, respectively. Note that \change{the horizontal stresses in $P_H$ and $P_V$}{$P_H$ and $P_V$ differs from those defined in \protect{\citep{verma_interaction_2022}}, and their horizontal components} have additional asymmetric though minor stress contribution directly from background mesoscale forcing $\vec{U}$, and that shear components have contributions from both along-front mean $\vec{u}^a$ and submesoscale deviation $\vec{u}^s$. And by comparing the pattern of $P'$ to those of $P_H$ and $P_V$, we can examine if the along-front mean shear or the submesoscale shear are dominant, which is a preferred choice when further decompositions become computationally challenging. The analysis below demonstrates that $P'$ and $P_H$ represent virtually the same submesoscale-BLT transfer in opposite directions, and that $P_V$ has significant contributions from both surface-intensified Ekman and frontal geostrophic shears. Although the direct contribution of mesoscale forcing $\vec{U}$ mainly appears in the total velocity components, its significance can accumulate steadily in time and space. Here the geostrophic shear production is 
\begin{equation}
    P_{Vg}=\frac{\overline{u'w'}}{f}\frac{\p \overline{b}}{\p y}-\frac{\overline{v'w'}}{f}\frac{\p \overline{b}}{\p x},
\end{equation}
and we can also examine similar geostrophic components in $P^a$ and $P^s$ to clarify generation or destruction of $\ske$ through the mean flow and self-interaction. Moreover, we do not include dissipation terms in \eqref{eq:ske} and \eqref{eq:tke} since the model only implements implicit numerical dissipation through the advection scheme \citep{silvestri_new_2024}.

\begin{figure}[pt]
  \centering
  \includegraphics[width=0.9\linewidth]{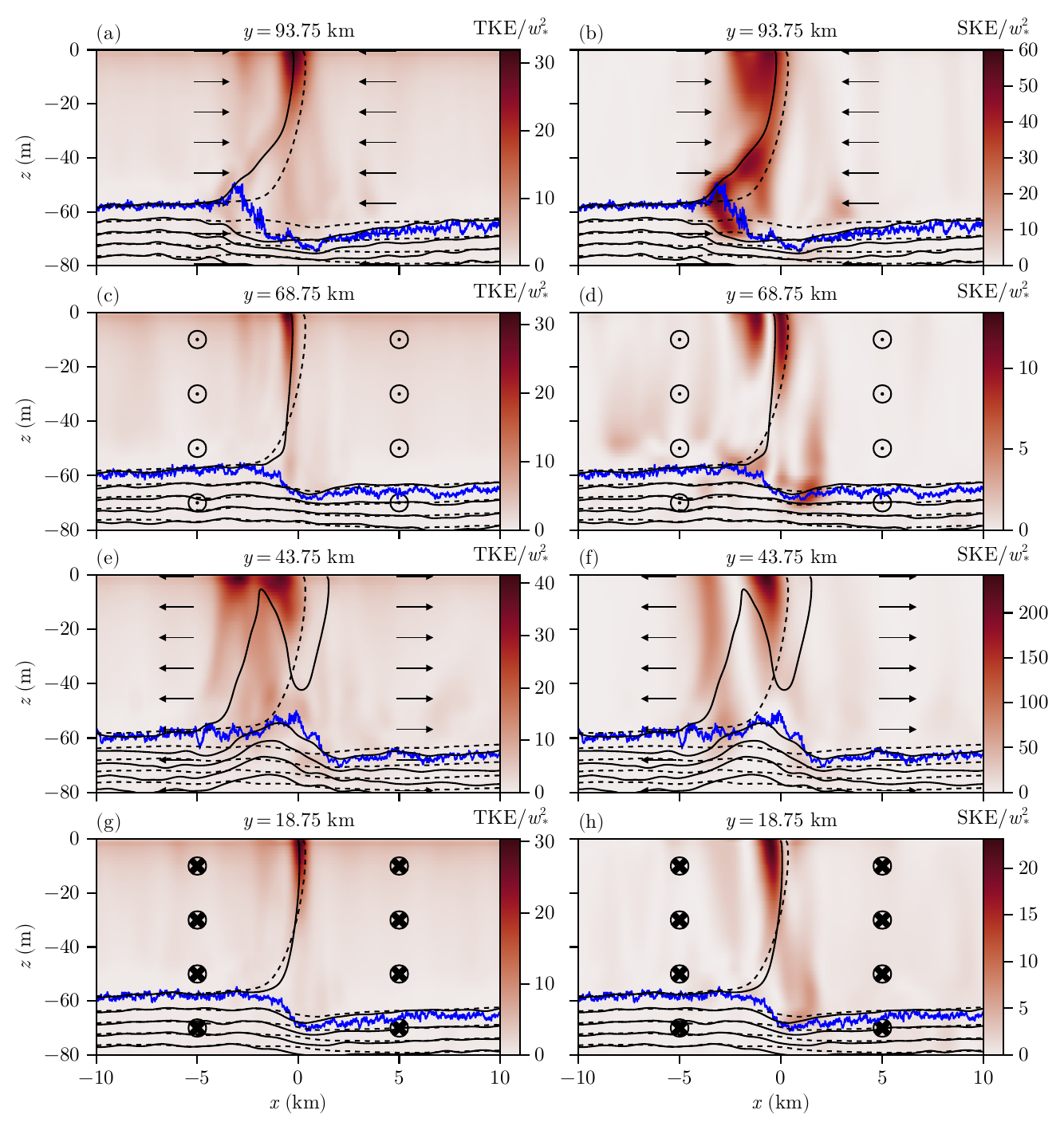}
\caption{TKE and SKE at 30~h along $y=\qtylist{93.75;68.75;43.75;18.75}{\kilo\meter}$ following the mean frontal flow (Fig.~\ref{fig:affields}b). Note that scales of the colorbars differ with locations and between TKE and SKE. Arrows indicate velocity vectors of the background mesoscale eddy forcing.\add[SP]{ Dashed black lines show isotherms of the along-front mean temperature, solid black lines for local isotherms, and blue lines denote the mixed layer.}}
  \label{fig:TKESKE}
\end{figure}

\begin{figure}[pt]
  \centering
  \includegraphics[width=0.9\linewidth]{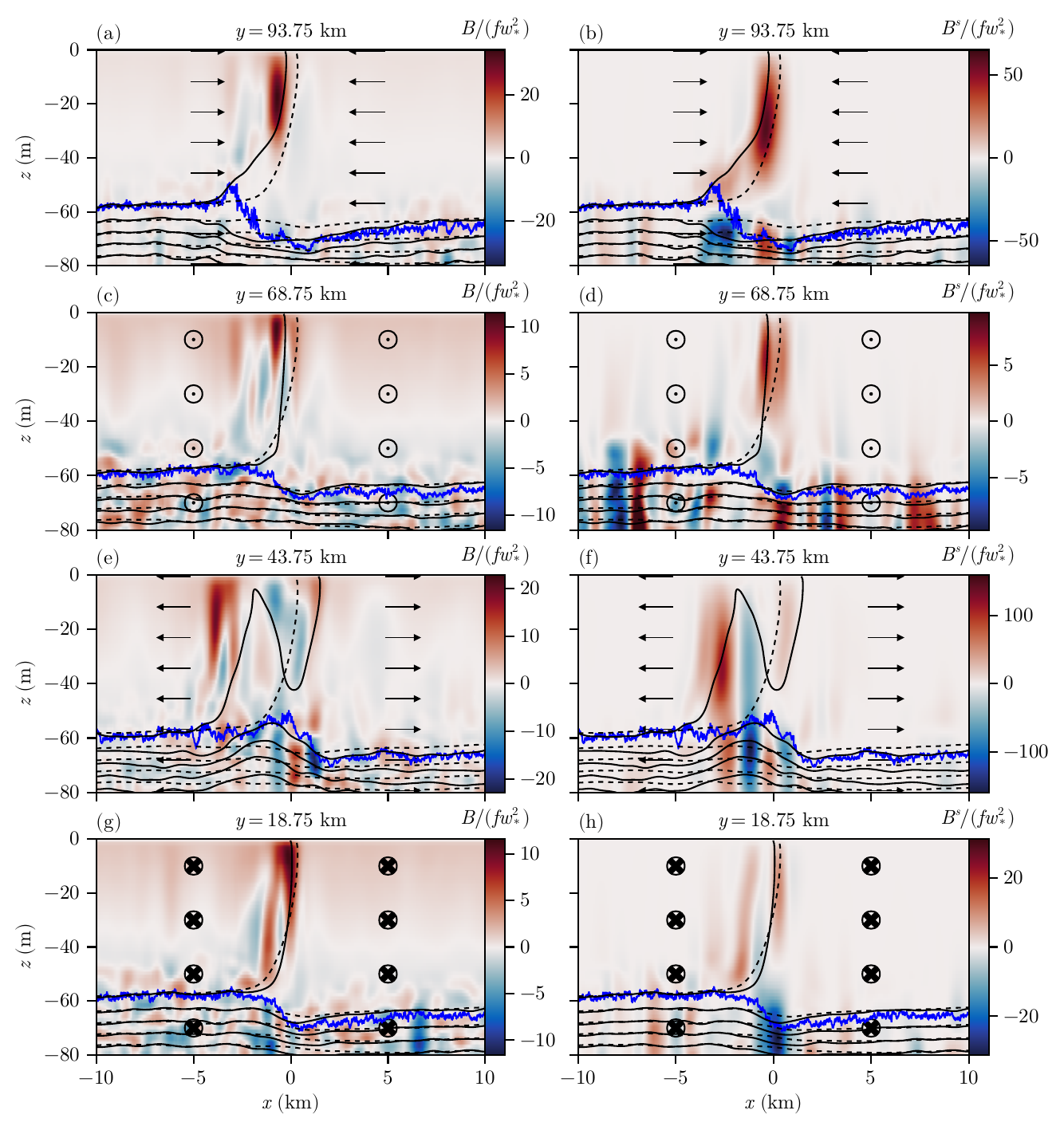}
\caption{TKE buoyancy production $B$ and SKE buoyancy production $B^s$ at 30~h along $y=\qtylist{93.75;68.75;43.75;18.75}{\kilo\meter}$ following the mean frontal flow (Fig.~\ref{fig:affields}b). Note that scales of the colorbars differ with locations and between $B$ and $B^s$. Arrows indicate velocity vectors of the background mesoscale eddy forcing.}
  \label{fig:BBs}
\end{figure}

\begin{figure}[pt]
  \centering
  \includegraphics[width=\linewidth]{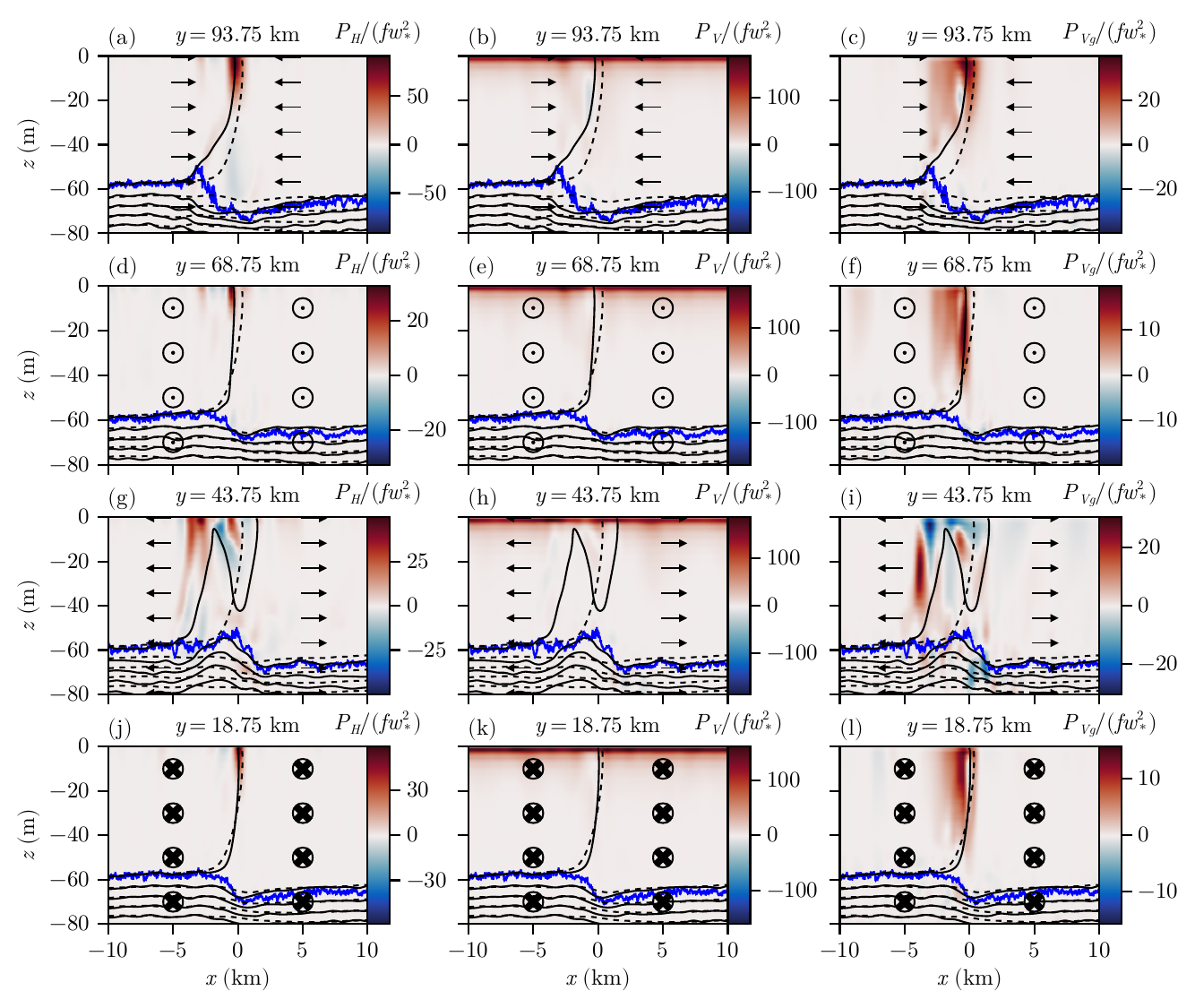}
\caption{TKE horizontal production $P_H$, vertical production $P_V$, and geostrophic vertical production $P_{Vg}$ at 30~h along $y=\qtylist{93.75;68.75;43.75;18.75}{\kilo\meter}$ following the mean frontal flow (Fig.~\ref{fig:affields}b). Arrows indicate velocity vectors of the background mesoscale eddy forcing. }
  \label{fig:PhPv}
\end{figure}

\begin{figure}[pt]
  \centering
  \includegraphics[width=\linewidth]{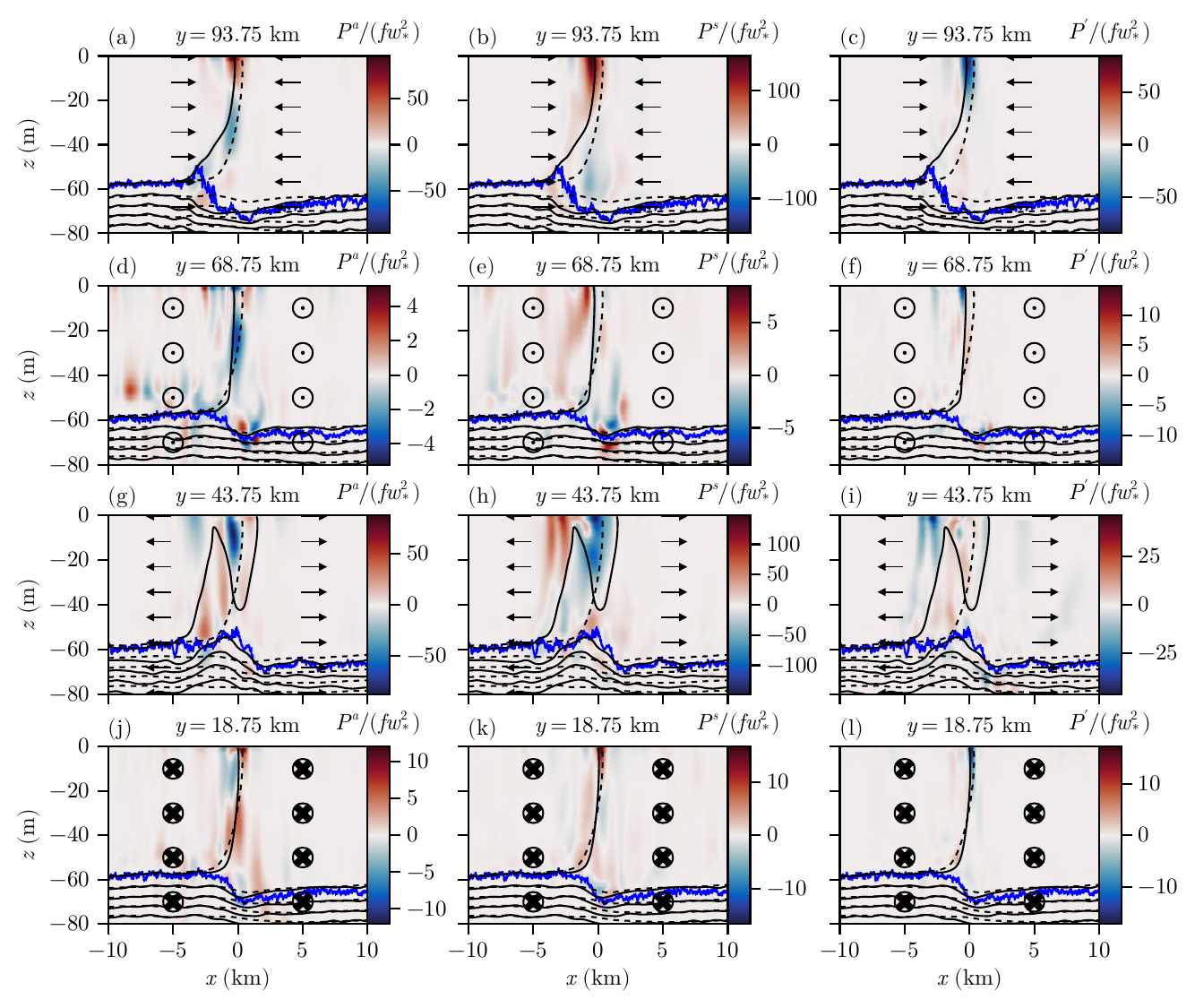}
\caption{SKE mean-flow production $P^a$, submesoscale and turbulent production $P^s$, $P'$ at 30~h along $y=\qtylist{93.75;68.75;43.75;18.75}{\kilo\meter}$ following the mean frontal flow (Fig.~\ref{fig:affields}b). Arrows indicate velocity vectors of the background mesoscale eddy forcing. }
  \label{fig:Past}
\end{figure}

Four vertical slices show drastically different budget contributions, especially right downstream of the strain extrema (Figs.~\ref{fig:TKESKE}-\ref{fig:Past}). Specifically, we follow the mean frontal flow direction and choose $y=\qtylist{93.75;68.75;43.75;18.75}{\kilo\meter}$. The strain is converging and diverging at $y=\qtylist{93.75;43.75}{\kilo\meter}$, respectively, while background eddy flow is augmenting and opposing the frontal flow respectively at $y=\qtylist{68.75;18.75}{\kilo\meter}$. The choice of slightly downstream locations from the strain extrema is supported by the advection of frontal flow in the negative $y$ direction, consistent with patterns on the horizontal slices of TKE and SKE (Fig.~\ref{fig:ufront30h}cd). These slices reveal how the large-scale flow modulates the local stratification, the distribution of kinetic energy, and the dominant terms in the turbulent and submesoscale energy budgets. 

The stratification at each slice deviates significantly from the along-front average, particularly in regions under extreme mesoscale strains (Figs.~\ref{fig:TKESKE}-\ref{fig:Past} solid vs. dash contours). Specifically, in the strong convergence region (e.g., at $y=\qty{93.75}{\kilo\meter}$), the frontal isotherm exhibits a slope close to that of the along-front mean isotherm above $z=\qty{-30}{\meter}$, but its slope below appears flatter, \change[SP]{indicative of submesoscale restratification near the edge of local SKE extreme}{consistent with local isopycnal flattening near the edge of local SKE extreme} (Figs.~\ref{fig:ufront30h}d). Isotherms around $x=0$ in the transition layer below $z=\qty{-60}{\meter}$ are displaced deeper compared with their along-front mean counterparts, suggesting mesoscale convergence-driven downwelling. In contrast, in the region under strong divergence ($y=\qty{43.75}{\kilo\meter}$), the frontal isotherm is dramatically distorted, showing significant curvature and heaving that implies more remarkable submesoscale stirring near the centre of local SKE extreme (Figs.~\ref{fig:ufront30h}d). Here, isotherms in the transition layer are shallower than those in the along-front average, consistent with divergent strain and submesoscale-induced upwelling. The other two slices ($y=\qtylist{18.75;68.75}{\kilo\meter}$), located in regions of weaker strain, show a frontal structure better resembles the along-front average, though with steeper slopes, perhaps due to locally dominant mixing that deepens the ML. \add[SP]{We therefore use these slice-based stratification signatures as qualitative contextual evidence, and do not pursue a dedicated restratification diagnosis from vertical slices in this manuscript.}

The vertical distributions of TKE and SKE reflect these differing dynamical regimes (Fig.~\ref{fig:TKESKE}). At the convergent-influenced slice ($y=\qty{93.75}{\kilo\meter}$), TKE is confined to the near-surface frontal isotherm, whereas SKE is about twice as large in both magnitude maximum and spatial extent, extending more to the left of the frontal isotherm and throughout the depth of the mixed layer (Fig.~\ref{fig:TKESKE}ab). This contrast consistently reveals that frontogenesis could be dominant above $z=\qty{-30}{\meter}$ where both TKE and SKE are strong, while submesoscale restratification controls the deeper frontal isotherm where only SKE is significant. Further downstream of the convergence zone ($y=\qty{68.75}{\kilo\meter}$), TKE is more tightly confined in $x$ with similar magnitude, likely due to cumulative strain impact, while SKE maximum is dramatically reduced to just half of TKE maximum (Fig.~\ref{fig:TKESKE}cd). The frontal isotherm splits SKE into a bimodal pattern above $z=\qty{-40}{\meter}$, likely due to differences between the local meandering front and its along-$y$ average. The most complex pattern appears slightly downstream of the divergent zone ($y=\qty{43.75}{\kilo\meter}$), where the strong curvature of the isotherm organizes both TKE and SKE into a bimodal structure (Fig.~\ref{fig:TKESKE}ef). TKE remains surface-intensified, while the two SKE modes are vertically distinct and extend to the two lower turning points of the frontal isotherm. Two TKE modes have similar maximum slightly higher than other slices, but SKE is dominant by the right mode with a maximum about four times that along $y=\qty{93.75}{\kilo\meter}$. We confirm later that this bimodal structure roughly corresponds to opposite energy transfer between TKE and SKE based on budget analysis. In regions of weaker strain ($y=\qty{18.75}{\kilo\meter}$), TKE is concentrated similar to that along $y=\qty{68.75}{\kilo\meter}$, while SKE is comparable in strength and confined mainly to one side of the isotherm (Fig.~\ref{fig:TKESKE}gh). To clarify the evolution of these heterogeneous across-front structures, we analyze terms in energy budgets of TKE and SKE.

The vertical buoyancy flux, a key energy conversion term, is most active near the strain extrema (Fig.~\ref{fig:BBs}). Near the converging zone ($y=\qty{93.75}{\kilo\meter}$), the TKE vertical buoyancy flux $B$ is positively concentrated around the frontal isotherm above $z=\qty{-40}{\meter}$ (Fig.~\ref{fig:BBs}a), indicating transfer to TKE from PE reservoir charged by surface cooling and Ekman-induced buoyancy flux \citep[e.g.,][]{thomas_destruction_2005, skyllingstad_baroclinic_2017,verma_interaction_2022}. The dominant mode of the SKE buoyancy flux $B^s$ is wider, twice stronger, centered around $z=\qty{-30}{\meter}$, and reaching to the base of the ML (Fig.~\ref{fig:BBs}b), representative of impact from submesoscale BI \citep{fox-kemper_parameterization_2008, johnson_modification_2024}. The SKE flux $B^s$ also shows fluctuating components in the transition layer, highlighting the more complex submesoscale processes other than BI like frontal overturning. These fluxes are weaker further downstream along $y=\qty{68.75}{\kilo\meter}$ with comparable spatial extent (Fig.~\ref{fig:BBs}cd). In particular, $B^s$ appears to be an order of magnitude smaller than its upstream counterpart and comparable to the collocated $B$ in terms of maximum. 

At the divergence-influenced slice ($y=\qty{43.75}{\kilo\meter}$), both buoyancy fluxes show a positive-negative dipole around the up-and-down frontal isotherm (Fig.~\ref{fig:BBs}ef). While the dipole for $B$ is confined near the surface and split by the tip of the curved isotherm, the $B^s$ dipole is located along and below the isotherm, extending even below the ML base \citep{siegelman_enhanced_2020}. The $B^s$ maximum magnitude here is more than twice that along $y=\qty{93.75}{\kilo\meter}$ and controls the deeper SKE pattern left of its maximum (Fig.~\ref{fig:TKESKE}f), yet the magnitude maximum of $B$ is lower than that in the converging zone. \add[SP]{Locally, resolved $B$ can exceed the domain-mean surface buoyancy-flux scale $B_0$ because frontal convergence and isopycnal tilting concentrate conversion into narrow regions. In this interpretation, Ekman buoyancy forcing provides part of the coupled frontal forcing context, while $B$ and shear-production terms co-occur as concurrent, spatially heterogeneous pathways rather than mutually exclusive regimes.} The vertical buoyancy fluxes along $y=\qty{18.75}{\kilo\meter}$ shows more vertical extent for $B$ yet weak ML component for $B^s$. The magnitude maximum for $B$ is comparable to that along $y=\qty{68.75}{\kilo\meter}$ with yet wider extent, while a strong negative flux in $B^s$ appears in the transition layer at $x=0$ and more than double the magnitude maximum of that along $y=\qty{68.75}{\kilo\meter}$. These drastically diverse patterns highlight how background mean flow, submesoscales and BLT interact closely to affect the buoyancy fluxes.

For the TKE budget specifically, the vertical buoyancy flux is less dominant than shear production terms (Fig.~\ref{fig:PhPv}).  Generally, horizontal shear production $P_H$ is positively concentrated at the frontal isotherm, consistent with energy transfer due to frontogenesis, while vertical shear production $P_V$ is dominated by near-surface Ekman shear except the front-concentrated geostrophic contribution. In particular, at $y=\qty{93.75}{\kilo\meter}$, the $P_H$ pattern resembles that of SKE and that of BLT-to-SKE transfer $P'$ with a sign-flip (Fig.~\ref{fig:PhPv}a,~\ref{fig:TKESKE}a,~\ref{fig:Past}a), consistent with downscale energy transfer under frontogenesis \citep{srinivasan_forward_2023}. That for $P_V$ is near-uniformly positive in the surface Ekman layer above $z=\qty{-10}{\meter}$ (Fig.~\ref{fig:PhPv}b), while its geostrophic component $P_{Vg}$ is dominantly positive around the frontal isotherm (Fig.~\ref{fig:PhPv}c), indicative of possible presence of symmetric instabilities \citep[e.g.,][]{thomas_symmetric_2013,skyllingstad_baroclinic_2017,wenegrat_current_2023}. At $y=\qty{68.75}{\kilo\meter}$, $P_H$ has negligible horizontal extent, and $P_V$ remains Ekman-dominated with similar front-concentrated geostrophic contribution (Fig.~\ref{fig:PhPv}def). At $y=\qty{43.75}{\kilo\meter}$, $P_H$ is positive to the negative-$x$ side of the frontal isotherm and turns into a dipole just right to the isotherm tip, while the geostrophic component of $P_V$ shows two dipoles at both locations (Fig.~\ref{fig:PhPv}gi). This complex pattern again demonstrate intricate coupling between BLT and submesoscale BI on the shear structure, regardless of the near-identical pattern in the full $P_V$ (Fig.~\ref{fig:PhPv}i). At $y=\qty{18.75}{\kilo\meter}$, the flux patterns resemble those along $y=\qty{68.75}{\kilo\meter}$ (Fig.~\ref{fig:PhPv}jkl).

The SKE budget further highlights the impact of large-scale mean flow and self-generation (Fig.~\ref{fig:Past}). Near the convergence zone ($y=\qty{93.75}{\kilo\meter}$), the budget is dominated by the submesoscale self-production $P^s$ near the surface, indicating local frontogenetic sharpening (Fig.~\ref{fig:Past}ab). Conversely, near the divergence zone ($y=\qty{43.75}{\kilo\meter}$), both the production from the along-front mean flow $P^a$ and the self-production term contribute significantly negative, creating a complex pattern of energy extraction and redistribution around the curved frontal isotherm (Fig.~\ref{fig:Past}gh). Across slices away from the strain extrema, all SKE production terms are comparatively weak. Geostrophic components in both $P^a$ and $P^s$ are generally concentrated around mid-depth of the ML and less than half of the full productions in term of maximum (not shown), suggesting the significance of ageostrophic impact in the SKE budget. Patterns of BLT-extraction $P'$ are virtually identical to those of $P_H$ with a flip of sign. This detailed budget analysis implies that the inhomogeneous mesoscale eddy field does not merely influence the frontal structure, but actively selects and amplifies different energy budget contributions between SKE and TKE at different locations along the front.

\begin{figure}[pt]
  \centering
  \includegraphics[width=0.9\linewidth]{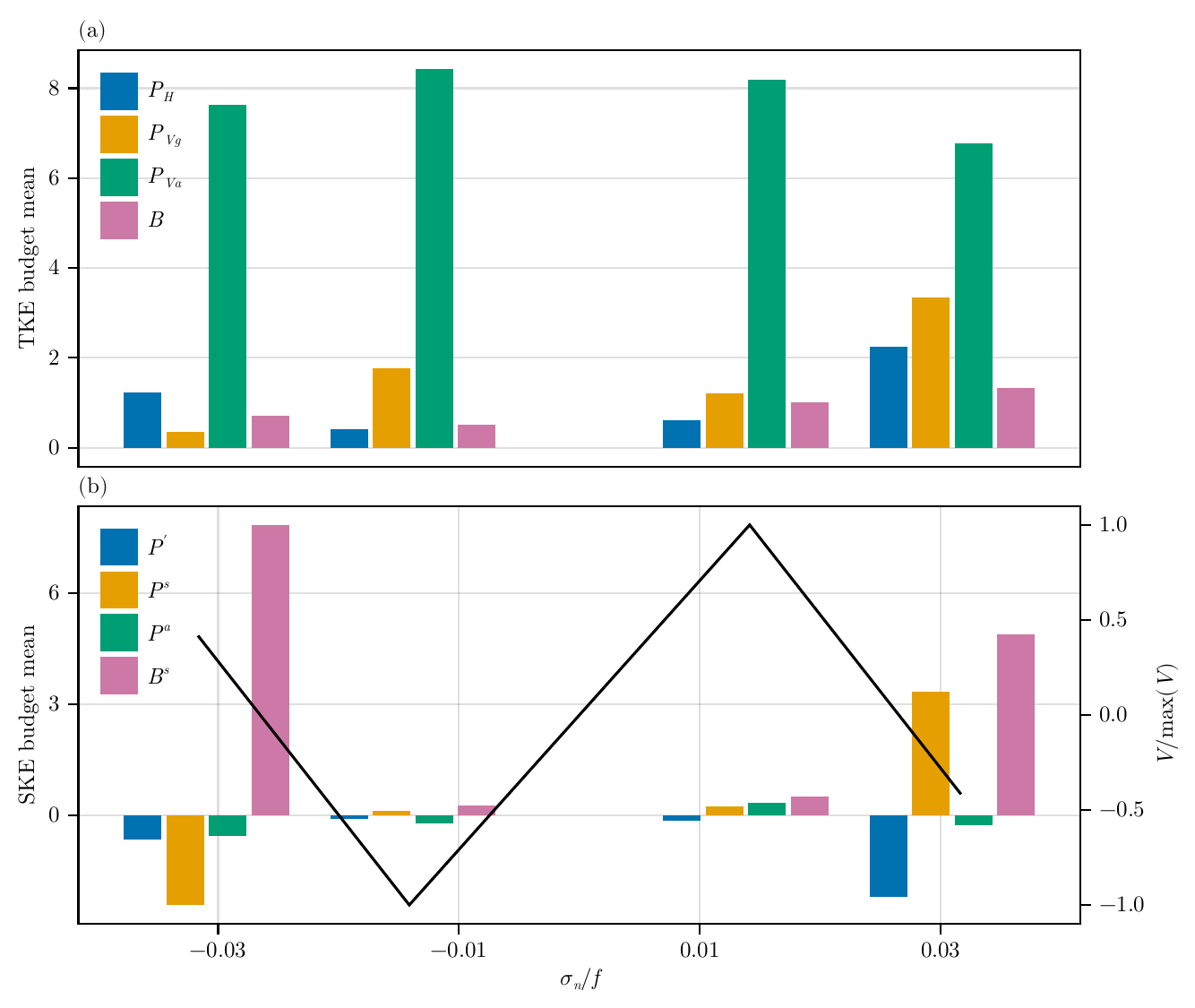}
\caption{Aggregated barplots of mean (a) TKE productions, and (b) SKE productions at four slices with different mesoscale strain $\sigma_n/f$. The spatial average is computed for $x\in [\qty{-4}{\kilo\meter}, \qty{4}{\kilo\meter}]$ and above the local MLD. \add[SP]{All production terms use the same normalization as a cross-front mean for $x\in [\qty{-4}{\kilo\meter}, \qty{4}{\kilo\meter}]$ and above the local MLD, so magnitudes can be compared directly across slices within each panel.} The black curve in (b) shows the normalized mesoscale velocity in $y$ at the four slices.}
  \label{fig:Pbarsmn}
\end{figure}

An aggregated barplot clarify these dramatic energy budget changes with the mesoscale background (Fig.~\ref{fig:Pbarsmn}). Specifically, we use a cross-front mean for $x\in [\qty{-4}{\kilo\meter}, \qty{4}{\kilo\meter}]$ and above the local MLD to aggregate various contributions along each vertical slice. For TKE, the mean production is dominated by the vertical ageostrophic shear due to BLT $P_{Va}=P_V-P_{Vg}$, while the geostrophic shear production can differ by about an order of magnitude between slices located in the most negative and positive strain regions, with normalized mean $P_{Vg}$ smaller than 0.35 and greater than 3.3, respectively (Fig.~\ref{fig:Pbarsmn}a). The mean horizontal shear production $P_H$ is stronger than that of $P_{Vg}$ only with the most negative strain. The mean buoyancy flux $B$ is generally moderate across slices, regardless of its totally different spatial pattern (Fig.~\ref{fig:BBs} left column). For SKE, the aggregation illustrates a correlation between strong productions and strong strains, similar to that between strain and SKE itself (Fig.~\ref{fig:Pbarsmn}b). In particular, the vertical buoyancy flux $B^s$ dominate the production for all slices and is the largest and only production term in the diverging strain region. The submesoscale self-production $P^s$ is the strongest destruction term here but the second largest production in the converging strain region. The BLT-destruction $P'$ contributes to $P_H$ and has a similar pattern as that for $P_H$ with a negative sign as expected. The average-flow production $P^a$ is mild and negative at three out of four slices. All terms are weak on slices with moderate normal  strains yet strongest meridional advection $V=\pm \max(V)$.  An alternative aggregation using productions at the TKE or SKE maxima shows a distinct pattern and much stronger across-slice variations and on-slice values, confirming the local spatial heterogeneity across the front (Fig.~\ref{fig:Pbarsmx}).

\remove[SP]{Former Section 5.3 (secondary mesoscale--Ekman interpretation) is removed from the main results section.}

\section{Discussion and conclusions}
\label{sec:discussion}

This study presents, to our knowledge, the first Large Eddy Simulation (LES) framework to resolve submesoscale and boundary layer turbulence (BLT) within a persistent, spatially heterogeneous mesoscale field. The primary finding is that mesoscale heterogeneity induces a first-order control on the evolution, structure, and energetics of an unstable front and its associated turbulence. By moving beyond the classical uniform-strain paradigm, we demonstrate that the background mesoscale field actively organizes submesoscale and BLT interactions, creating a spatial pattern of turbulent hotspots whose locations are tied to the underlying large-scale flow. \add[SP]{Note that our goal is to diagnose correlation within a single, dynamically consistent simulation rather than a clean causal separation.}

Our model setup integrates a prescribed mesoscale forcing with a two-sided mixed layer front initially in geostrophic balance \citep{thomas_friciton_2008, hamlington_langmuir_2014}. Thanks to the computational power of distributed GPUs and the Oceananigans configuration \citep{ramadhan_oceananigans_2020}, our model is able to resolve BLT under the mesoscale forcing. The excitation of BLT by surface wind stress and cooling is fast, and the front with along-front wind component experiences destabilizing Ekman buoyancy flux that advects cold over warm waters \citep[e.g.,][]{thomas_destruction_2005, wenegrat_current_2023}, developing significant along-front variabilities of front width, curvature, mixed-layer depth (MLD), and other frontal properties under mesoscale forcing. \change[SP]{These}{By analyzing vorticity evolution, evolution of along-front statistics, and along-front heterogeneity, we confirm that these} variations correlate strongly with the background mesoscale, in particular for which convergence (divergence) reduces(increases) the frontal width yet deepens(shoals) the MLD.

A triple flow decomposition allowed us to quantify the kinetic energy budgets for BLT (TKE) and submesoscales (SKE) and examine their spatial variability across distinct dynamical regimes. We observed that extreme mesoscale strain colocates with both amplified production and destruction terms, and strong divergence further coupled to significant cross-sectional variability. Specifically, under strong mesoscale convergence, TKE production via horizontal and vertical geostrophic shear is maximized. Here, the mean geostrophic shear production reaches approximately half the magnitude of the ageostrophic term, a ratio consistent with the presence of a strong submesoscale source, though ageostrophic production remains dominant \citep{dong_submesoscales_2024}. Similarly, SKE self-production and destruction by BLT are enhanced in convergent zones. Conversely, under strong divergence, the self-production term becomes negative, acting as a sink, while the vertical buoyancy flux remains as the dominant source for SKE, highlighting its critical role in sustaining submesoscale instabilities such as baroclinic instability.

\add[SP]{Returning to the three hypotheses posed in the Introduction, the analysis supports a clear interpretation. First, the hypothesis that mesoscale convergence sharpens the front and enhances geostrophic-shear-driven TKE is supported by the along-front narrowing of the frontal width, the deeper MLD in convergent sectors, and the concentration of both horizontal and geostrophic vertical shear production near the frontal isotherm in the strongly convergent slice. Second, the hypothesis that mesoscale divergence favors frontal distortion and buoyancy-driven SKE pathways is also supported: the divergent sector shows the strongest frontal curvature and heaving, the most spatially complex SKE structure, and a dominant $B^s$ signal together with negative self-production, consistent with buoyancy conversion playing the leading role there. Third, the hypothesis that along-front mesoscale heterogeneity modulates the timing and expression of local MLI growth is supported more indirectly. We do not claim a strict causal delay law from this single simulation, but the spatial progression from relatively coherent frontal sharpening in convergent sectors to stronger meandering, distortion, and energetically distinct instability signatures in weaker-strain and divergent sectors shows that the background mesoscale field organizes where and how local instability pathways emerge.}

These findings offer critical context for interpreting previous process studies. While prior high-resolution LES utilizing zero or uniform background strain have focused on budget analysis under volume-wise or along-front averaging \citep{sullivan_frontogenesis_2018, verma_interaction_2022, bodner_modifying_2023, johnson_modification_2024}, our results clarify that the multiscale energy budget contributions can be spatially intermittent under heterogeneous mesoscale forcing. The simultaneous coexistence of these opposing yet connected dynamical regimes requires more careful budget formulations than those in strain-free or uniform-forcing models. Furthermore, compared to hydrostatic regional models which capture mesoscale to submesoscale variability but parameterize BLT \citep[e.g.,][]{srinivasan_forward_2023}, our explicit resolution of non-hydrostatic BLT reveals order-of-magnitude along-front variability in TKE fluxes with distinct spatial patterns under larger-scale modulation. This suggests that hydrostatic models may systematically underestimate certain variabilities of vertical mixing in frontal regions due to lack of enough large-scale coupling in current subgrid-scale parameterizations.

These results also have significant implications for the representation of upper ocean dynamics in large-scale models. The pronounced along-front heterogeneity in TKE and SKE implies that vertical transport of heat, carbon, and other tracers is likewise highly localized \citep{sinha_submesoscales_2023, mahadevan_impact_2016, taylor_submesoscale_2023, ferrari_frontal_2011, su_ocean_2018}. Our budget analysis confirms that domain-integrated vertical fluxes are likely dominated by these distinct hotspots. Moreover, the results suggest that the mesoscale field, along with surface forcing, could selectively trigger submesoscale instabilities via secondary perturbations, rather than allowing broad, random growth along the front. This indicates a dynamic coupling between mesoscale-driven and instability-driven submesoscale processes that have conventionally been treated in isolation. The drastic variability in budget fluxes implies that current climate and general circulation models likely miss crucial, spatially intermittent interactions \citep{bachman_parameterization_2017,bodner_modifying_2023}. These insights were only possible due to the unique combination of a $\qty{100}{\kilo\meter}$-domain, which holds the mesoscale, and meter-scale grid resolution, which resolves the non-hydrostatic dynamics. 

\change[SP]{We acknowledge certain idealizations necessary to isolate the target physics. The upper limit on resolved scales is set by the $\qty{100}{\kilo\meter}$ domain, and the imposed mesoscale forcing follows a controlled quadrupole pattern without an eddy-driven surface temperature signature. The absence of feedback terms to the mesoscale flow in the momentum equations also breaks the clarity in formulating derivative budgets like those for vorticity, since the derivation will involve isolated horizontal gradients of $\vec{U}$. In the vertical, the $\qty{250}{\meter}$ depth and the absence of a full thermocline constrain interactions with the deep ocean. Finally, the integration time captures the transient adjustment of the front rather than a full mesoscale-submesoscale equilibrium. In future studies we will report on the extended integration time of the simulation when turbulence is fully developed. These choices, however, are what define this as a controlled numerical experiment, allowing us to isolate the specific mechanism of mesoscale-submesoscale-BLT coupling from the full complexity of a realistic ocean energy cascade. The sensitivity of quantitative correlations to model setup and initialization remains an important area for investigation\protect{\citep{atkinson_near_2025}}.}{We acknowledge several idealizations that define the scope of this process study. The upper limit on resolved scales is set by the $\qty{100}{\kilo\meter}$ domain, and the imposed mesoscale forcing follows a controlled stationary quadrupole pattern without an eddy-driven surface temperature signature. Our one-way-coupled formulation prescribes $\vec{U}$ and does not include two-way perturbation feedback onto a prognostic mesoscale state; therefore, mechanisms that depend on fully interactive mesoscale evolution are outside the model scope. In the vertical, the $\qty{250}{\meter}$ depth and the absence of a full thermocline constrain interactions with the deep ocean. Note that these two points are dynamically coupled: we did try first with a more realistic freely evolving mesoscale field. The issue was that we weren't able to maintain the  mesoscale eddies given the shallow depth of the domain and prescribed density structure. They dissipated too fast and thus were not able to provide the necessary forcing for the  submesoscale. So the current setup was not an oversight or a simplification but rather it was necessary to capture the physics in this problem of interest. Moreover, the integration time captures the transient adjustment of the front rather than a full mesoscale-submesoscale equilibrium. Finally, a more direct approach to isolate the impact of mesoscale straining on submesoscale would be to compare solutions with and without mesoscale forcing, but it is not feasible to run an entire simulation without mesoscale forcing due to required resources. Also, that has already been done in a smaller domain, although its earlier transient evolution is less explored and dynamically intriguing analysis focuses on outputs after \protect{\qty{10}{\day} \citep{hamlington_langmuir_2014,  johnson_modification_2024}}. In future studies we will report on extended integrations and reruns with a time-evolving mesoscale field to assess robustness and generality. These choices nonetheless define a controlled numerical experiment that isolates mesoscale-submesoscale-BLT coupling pathways under prescribed heterogeneous forcing. The sensitivity of quantitative correlations to model setup and initialization remains an important area for investigation\protect{\citep{atkinson_near_2025}}.}

\add[SP]{As a brief observational note, we retain only that asymmetric near-surface vorticity-band patterns co-occur with mesoscale structure and wind-forced flow, but we do not treat this as a primary mechanistic result in the present study. In particular, we avoid definitive attribution to vortex-Rossby-wave dynamics or other mechanisms that require a fully interactive mesoscale state beyond the scope of the present one-way-coupled setup. The main conclusions are based on the SKE/TKE budget diagnostics in the preceding budget subsection.}
\note[SP]{Detailed linear-wave/edge-wave interpretation is not included in the main text and can be provided in an appendix or supplement if requested.}

The versatility of this framework extends its utility beyond the current study. The explicit resolution of fine-scale frontal features provides a rich dataset for testing and refining subgrid-scale parameterizations for coarser-resolution ocean models. Our findings strongly suggest that such parameterizations must be scale-aware and, crucially, location-aware accounting for the local mesoscale strain and vorticity \citep{perezhogin_stable_2024,bodner_data_2025}. The model can also be adapted to investigate multiscale influences on tracer transport, such as the vertical exchange of heat and carbon, which are vital for climate and biogeochemical studies. These applications, accelerated by GPU computation, highlight the potential of this modeling approach to guide future research and bridge the gap between small-scale turbulence and large-scale ocean circulation.

In summary, this study presents an idealized yet powerful numerical framework for investigating heterogeneous frontal turbulence in the upper ocean mixed layer. We utilized a GPU-based LES to simulate the evolution of submesoscale and boundary layer turbulence under prescribed mesoscale field, surface wind forcing, and cooling, achieving a horizontal resolution of \qty{4.88}{\meter} in a domain of \qty{100}{\kilo\meter}. Compared to previous works focused on either submesoscale–BLT interactions or mesoscale–submesoscale energy transfers, our non-hydrostatic LES bridge these scales within a single simulation. Our results demonstrate that mesoscale inhomogeneity connects strongly to distinct structure and energetics of frontal turbulence. Future work can explore more realistic mesoscale representation, e.g. time-varying background velocity $\vec{U}$, or inclusion of surface waves.

\section*{Acknowledgements}{We are thankful to Raffaele Ferrari and Gregory LeClaire Wagner for insightful initial discussions of the model development. This research used resources of the National Energy Research Scientific Computing Center, a DOE Office of Science User Facility supported by the Office of Science of the U.S. Department of Energy under Contract No. DE-AC02-05CH11231 using NERSC award BER-ERCAP0033352.}
%
%
\section*{Declaration of interests}{The authors report no conflict of interest.}
%
\section*{Data availability statement}{The code that supports the findings of this study is openly available in \url{https://github.com/bodner-research-group/LESStudySetup.jl}, and the simulation data will also be openly available once properly processed. In accordance with the journal's policy, the authors declare the use of generative AI tools, specifically ChatGPT and Gemini, to assist with code debugging and the linguistic refinement of the text originally drafted by the authors. The authors have reviewed all AI-generated output and assume full responsibility for the accuracy and integrity of the final manuscript.}
%
%

\begin{appen}
\section{Imposed mesoscale eddy velocity parameterization}\label{app:eddyv}
We set the eddy velocity as a sum of four components defined in transformed coordinates centered at eddy centers. In particular, two warm-eddy centers are located at (\qty{25}{\kilo\meter}, \qty{25}{\kilo\meter}) and (\qty{75}{\kilo\meter}, \qty{75}{\kilo\meter}), while two cold-eddy centers are located at (\qty{25}{\kilo\meter}, \qty{75}{\kilo\meter}) and (\qty{75}{\kilo\meter}, \qty{25}{\kilo\meter}). The eddy radius is $R=L_x/4=\qty{25}{\kilo\meter}$. \add[SP]{This parameterization is intended as an idealized analytic forcing construction, not as a freely evolving exact mesoscale vortex equilibrium.} For each eddy-centered polar coordinates with distance $r$:
\begin{equation}
    \xi(r) = 2\frac{R - r}{\pi R L_e} - \pi\frac{L_e - 1}{2}\ ,
\end{equation}
where $L_e=0.9$ is a dimensionless eddy frontal width parameter. Then we define for a warm-core eddy a few variables dependent on $r$ and $\xi$ as:
\begin{align}
    u^B &= \frac{2 \Phi g r}{f R^2  \sigma^2}  \exp\left(-\frac{r}{ R^2  \sigma^2}\right)    \ ,\\
    \frac{\partial b}{\partial \xi} &= - \text{Int}(0 < \xi < \pi_0) g\alpha a \Delta T^e \Delta m\frac{\sin(\xi)^2 - \cos(\xi)^2 + 1}{ \pi}\ ,\\
    h_m &=\text{Int}(\xi > \pi_0) +\text{Int}(0 < \xi < \pi_0) \left[1 - \frac{\pi - \xi - \sin(\pi - \xi) \cos(\pi - \xi)}{ \pi}\right]\ ,
\end{align}
where $\Phi=\qty{0.01}{\meter}$ is the barotropic vortex surface amplitude, $\sigma^2=1$ is the dimensionless vortex spread, $\pi_0=3.1415926535897$, and $\Delta T^e = \qty{0.1}{\celsius}$ is the eddy temperature difference. Then the warm-eddy tangential velocity at polar angle $\varphi$ is:
\begin{equation}
    u^t(\varphi) = 
    \begin{cases}
    \frac{\partial \xi}{\partial r} \frac{\partial b}{\partial \xi} \frac{2}{3f} + u^B,& \text{if } z>-h=-(m_0 + h_m \Delta m )\\
    \frac{\partial \xi}{\partial r} \frac{\partial b}{\partial \xi} \frac{2(L_z + z)^3}{3f(L_z - h)^3}  + u^B,              & \text{otherwise}
    \end{cases}
\end{equation}
where $\partial \xi/\partial r=- 2\pi / (R  L_f)$,  $m_0=\qty{60}{\meter}$ is the initial mixed-layer depth, and $\Delta m=\qty{30}{\meter}$ is the mixed-layer depth difference scaled by $h_m$ between warm and cold eddies. \change[SP]{Note that the first term is a baroclinic adjustment that decrease to zero at the bottom of the domain.}{Note that the first term is a prescribed baroclinic adjustment, tapered toward zero at the bottom of the domain, chosen to keep the imposed buoyancy and velocity structure internally consistent with the intended thermal-wind scaling in this idealized setup.} The velocity components in the original coordinate are $u^t \sin (\varphi)$ and $-u^t \cos(\varphi)$. The velocities for cold-core eddies are similarly defined, with a different sign for $u^B$ and $\partial b/\partial \xi$ and $h=m_0-h_m \Delta m$. The sum of all four eddy velocity fields establishes a geostrophically balanced background flow $\vec{U}$. \change[SP]{The simulations use $\vec{U}$ only to advect horizontal velocities and temperature, and we assume that other advection terms related to $\vec{U}$ are negligible on the timescales and depth relevant to this setup.}{The simulations prescribe $\vec{U}$ as a stationary background that advects perturbation fields; reciprocal feedback from perturbations onto a prognostic mesoscale state is not represented in this formulation.}

\section{Imposed initial temperature and velocity}\label{app:initialTv}
We construct this setup by joining two profiles given by 
\begin{equation}
    \frac{M^2_0  L_f L_x }{180 \alpha g} \left[1-\tanh {\left(\frac{x}{L_f L_x/90}\right)}+\tanh {\left(\frac{x-L_x/2}{L_f L_x/90}\right)}\right]\left[\tanh \left(\frac{z+m_0}{\Delta m^f}\right)+1\right] + \Gamma_T
\end{equation}
for $x \leq L_x/2$, and 
\begin{equation}
    \frac{M^2_0  L_f L_x }{180 \alpha g} \left[\tanh {\left(\frac{x-L_x/2}{L_f L_x/90}\right)}-\tanh {\left(\frac{x-L_x}{L_f L_x/90}\right)}-1\right]\left[\tanh \left(\frac{z+m_0}{\Delta m^f}\right)+1\right] + \Gamma_T
\end{equation}
for $x > L_x/2$, where $L_f=0.9$ a dimensionless frontal width scale, $L_x=\qty{100}{\kilo\meter}$ the zonal domain size, and
\begin{equation}
    \Gamma_T=\frac{0.5}{\alpha g}\left\{\left(N^2_s+0.1 N^2_T\right)z+\Delta m^f \left[\left(N^2_s- N^2_T\right)\ln \left(\frac{\cosh \left(\frac{z+m_0}{\Delta m^f}\right)}{\cosh \left(\frac{m_0}{\Delta m^f}\right)}\right)+0.9 N^2_T \ln \left( \frac{\cosh \left(\frac{z+1.5m_0}{\Delta m^f}\right)}{\cosh \left(\frac{1.5m_0}{\Delta m^f}\right)}\right)\right]\right\}
\end{equation}

\section{Illustration of the zoom-in domain}\label{app:zoomin}
\begin{figure}[pt]
    \centering
    \includegraphics[width=\linewidth]{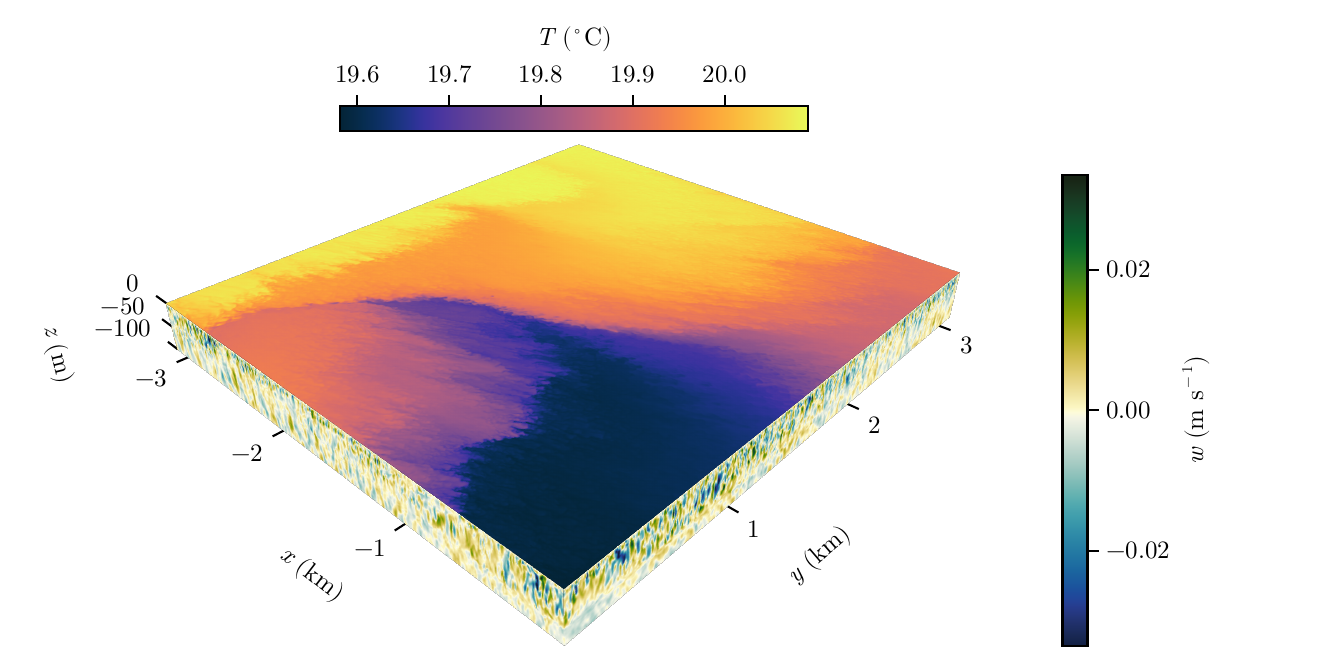}
    \caption{Visualization of surface temperature field $T$ and cross-section vertical velocity field $w$ in the zoom-in computational domain in Fig.~\ref{fig:Tlast} above $z=\qty{-100}{\meter}$ at \qty{7.5}{\day}.}
    \label{fig:Tlast_app}
\end{figure}

\section{Illustration of parameter choice}\label{app:mldass}
\begin{figure}[pt]
    \centering
    \includegraphics[width=0.9\linewidth]{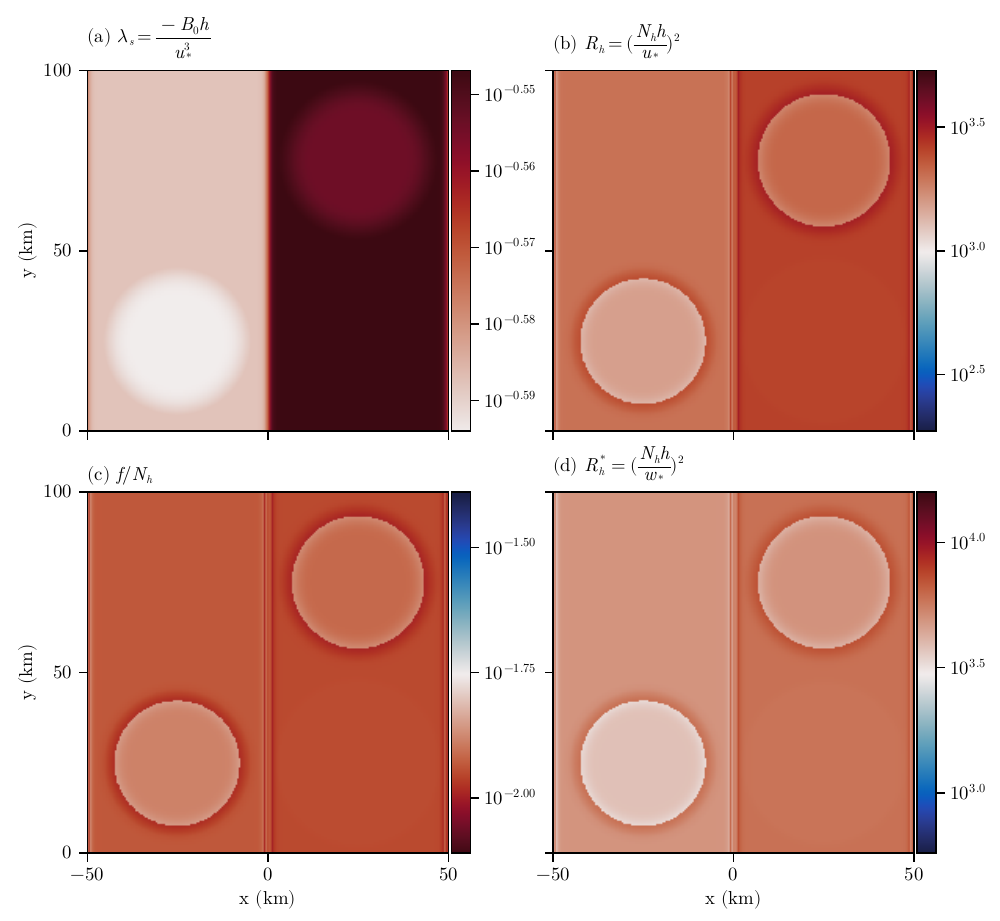}
    \caption{Initial parameter distributions for evolution of the mixed-layer depth following the framework of \citet{legay_framework_2024}. Shown are (a)~$\lambda_s=-B_0 h/u_*^3$ the relative contribution of the cooling and the wind, (b)~$R_h=(N_h h/u_*)^2$ and $R_h^*=(N_h h/w_*)^2$ the stability of the mixed layer relative to the wind, (c)~the importance of the Earth's rotation relative to the stratification, and (d)~$R_h^*=(N_h h/w_*)^2$ the stability of the mixed layer relative to the cooling.}
    \label{fig:mldass_d0}
\end{figure}

\section{Evolution of \add[SP]{horizontal }divergence}\label{app:div}
\begin{figure}[pt]
    \centering
    \includegraphics[width=0.9\linewidth]{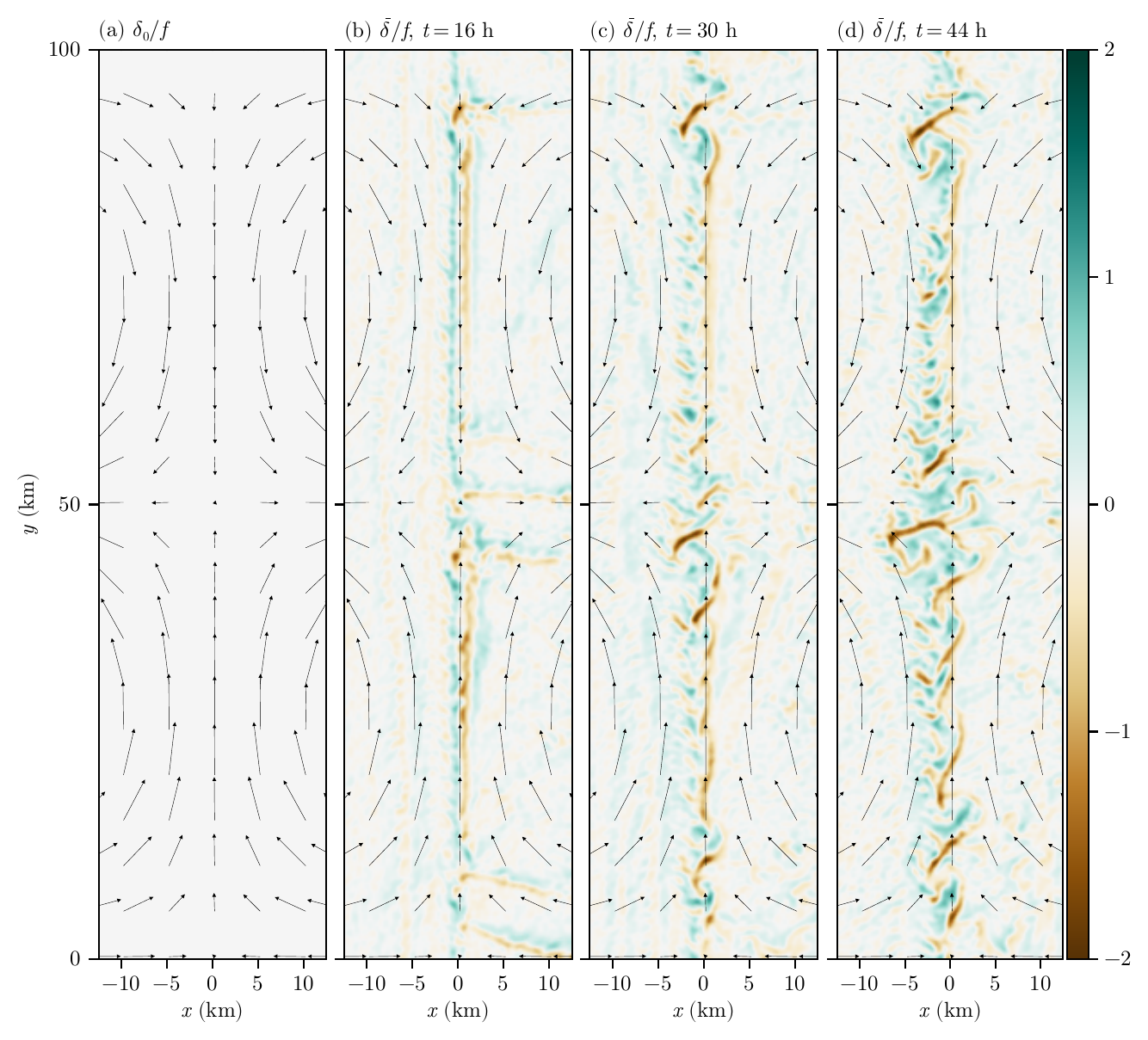}
    \caption{Snapshots of normalized surface \add[SP]{horizontal }divergence at (a) initialization, (b) 16~h, (c) 30~h, and (d) 44~h. The initial divergence in (a) is zero, while those in other panels are calculated on the meter-scale grid after a \qty{300}{\meter}-Gaussian kernel smoothing. Arrows indicate velocity vectors of the eddy forcing. }
    \label{fig:div}
\end{figure}

\section{Multiscale spectra profiles}\label{app:spec}
\begin{figure}[pt]
    \centering
    \includegraphics[width=0.9\linewidth]{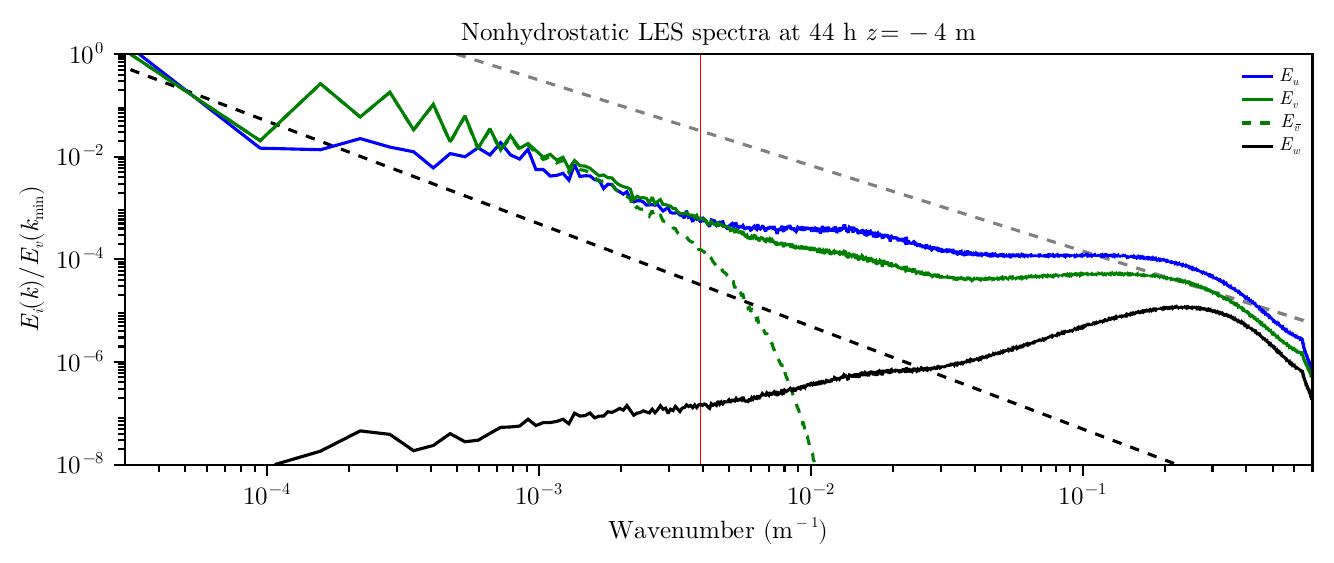}
    \caption{Circularly integrated spectra of the velocity vector $(u,v,w)$ normalized by values at the smallest wavenumber, examined at $z=\qty{-4}{\meter}$ depth. Dashed lines denote the $k^{-2}$ (black) and $k^{-5/3}$ (gray) slopes for reference. The red vertical line denotes the wavenumber $\sqrt{2 \ln{2}}L_{filter}^{-1}$ corresponding to the half-power cutoff wavenumber for a Gaussian kernel with $L_{filter}=\qty{300}{\meter}$ scale, used for separating boundary layer turbulence from larger scales, and dashed green line shows $E_v$ filtered with a Gaussian kernel at that scale. We calculate the 1D spectra from 2D Fast Fourier transforms by summing cospectra within the same 1D wavenumber bins. The 1D wavenumber for a cospectrum is the length of its 2D wavenumber vector. }
    \label{fig:spec}
\end{figure}

\section{Aggregated energy budgets at maximum TKE/SKE}\label{app:Pbarmx}
\begin{figure}[pt]
  \centering
  \includegraphics[width=0.9\linewidth]{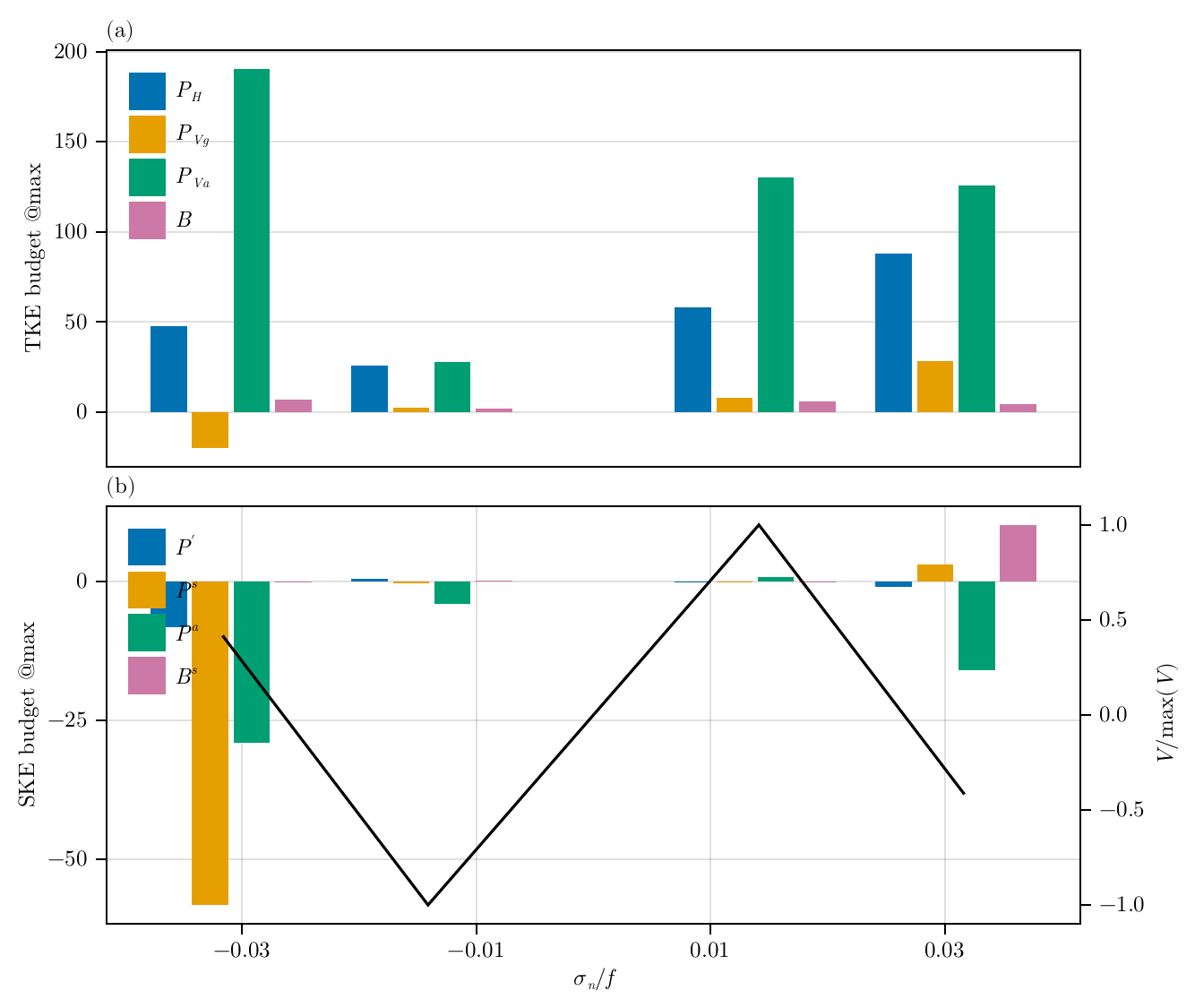}
\caption{Aggregated barplots of (a) TKE productions, and (b) SKE productions at maximum TKE/SKE indexes across four slices with different mesoscale strain $\sigma_n/f$. The black curve in (b) shows the normalized mesoscale velocity in $y$ at the four slices.}
  \label{fig:Pbarsmx}
\end{figure}

\remove[SP]{Appendix on a hydrostatic run referred in former Section 5.3 is removed}

\end{appen}\clearpage

\bibliographystyle{jfm}
\bibliography{jfm}

\end{document}